\documentclass[prd,superscriptaddress,twocolumn,showpacs,amsmath,amssymb]{revtex4}
\usepackage{amsmath}
\usepackage{graphicx}
\usepackage{color}
\usepackage{dcolumn}
\usepackage{bm}
\usepackage{slashed}
\usepackage{slashbox}
\usepackage{pdfpages}
\usepackage{epstopdf}
\usepackage{booktabs}
\usepackage{multirow}
\usepackage{float}
\usepackage[colorlinks,linkcolor=red,anchorcolor=green,citecolor=blue,CJKbookmarks=True]{hyperref}

\newcommand{\bea}{\begin{eqnarray*}}
\newcommand{\eea}{\end{eqnarray*}}
\newcommand{\bean}{\begin{eqnarray}}
\newcommand{\eean}{\end{eqnarray}}
\newcommand{\eqs}[1]{Eqs. (\ref{#1})}
\newcommand{\eq}[1]{Eq. (\ref{#1})}
\newcommand{\meq}[1]{(\ref{#1})}
\newcommand{\grad}{\nabla}

\newcommand{\non}{\nonumber \\}

\newcommand{\hsp}{\hspace{0.1mm}}

\newcommand{\pp}{\partial}

\begin{document}
\title{QNMs of Slowly Rotating Einstein-Bumblebee Black Hole}

\author{Wentao Liu}
\affiliation{Department of Physics, Key Laboratory of Low Dimensional Quantum Structures and Quantum Control of Ministry of Education, and Synergetic Innovation Center for Quantum Effects and Applications, Hunan Normal
University, Changsha, Hunan 410081, P. R. China}

\author{Xiongjun Fang}
\email[Corresponding author: ]{fangxj@hunnu.edu.cn} \affiliation{Department of Physics, Key Laboratory of Low Dimensional Quantum Structures and Quantum Control of Ministry of Education, and Synergetic Innovation Center for Quantum Effects and Applications, Hunan Normal
University, Changsha, Hunan 410081, P. R. China}

\author{Jiliang Jing}
\email{jljing@hunnu.edu.cn}\affiliation{Department of Physics, Key Laboratory of Low Dimensional Quantum Structures and Quantum Control of Ministry of Education, and Synergetic Innovation Center for Quantum Effects and Applications, Hunan Normal
University, Changsha, Hunan 410081, P. R. China}

\author{Jieci Wang}
\email{jcwang@hunnu.edu.cn}\affiliation{Department of Physics, Key Laboratory of Low Dimensional Quantum Structures and Quantum Control of Ministry of Education, and Synergetic Innovation Center for Quantum Effects and Applications, Hunan Normal
University, Changsha, Hunan 410081, P. R. China}

\begin{abstract}
We have studied the quasinormal modes (QNMs) of a slowly rotating black hole with Lorentz-violating parameter in Einstein-bumblebee gravity. We analyse the slow rotation approximation of the rotating black hole in the Einstein-bumblebee gravity, and obtain the master equations for scalar perturbation, vector perturbation and axial gravitational perturbation, respectively. Using the matrix method and the continuous fraction method, we numerically calculate the QNM frequencies. In particular, for scalar field, it shows that the QNMs up to the second order of rotation parameter have higher accuracy. The numerical results show that, for both scalar and vector fields, the Lorentz-violating parameter has a significant effect on the imaginary part of the QNM frequencies, while having a relatively smaller impact on the real part of the QNM frequencies. But for axial gravitational perturbation, the effect of increasing the Lorentz-violating parameter $\ell$ is similar to that of increasing the rotation parameter $\tilde{a}$.\\


\end{abstract}


\maketitle

\section{Introduction}\label{Sec.1}
Lorentz symmetry is a very basic and important spacetime symmetry, which is not only the basis of quantum field theory and the standard model of modern particle physics, but also plays a fundamental role in relativity. Motivated by establishing the quantum gravity, and the evidence from high energy cosmic rays, the possibility of Lorentz symmetry breaking (LSB) on Planck energy scale is expected and  has been extensively discussed \cite{Kostelecky1998}. The suggestions for sources of Lorentz violation (LV) included string field theory \cite{Kostelecky19891,Kostelecky1991}, loop quantum gravity theory \cite{Gambini1999}, noncommutative field theories \cite{Kostelecky2001,Ferrari2007}, quantum gravity inspired spacetime foam scenarios \cite{Garay1998}, and varying couplings \cite{Kostelecky20030,Bertolami2004}. However, in the low energy scale, this Lorentz symmetry breaking and its effects should be largely supressed, otherwise it would be inconsistent with many current ground-based experimental observations. Thus, over the past few decades, the search for quantum gravity effects in the low energy scale attached much attention. Especially, the standard-model extension is a candidate which offers a broad theoretical foundation for testing Lorentz symmetry.

The Lorentz symmetry is broken when a vector field has a nonzero vacuum expectation value. A straightforward effective theory of gravity with a standard model extension LV is the Bumblebee model, where the Lorentz violation results from the dynamics of a single vector or axial-vector field $B_\mu$, also known as the bumblebee field. Kostelecky and Samuel first used the bumblebee gravitational model to investigate the effects of spontaneous LV \cite{Kostelecky1989, Gomes2010}. Bertolami and P\'{a}ramos derived the vacuum solutions of this gravity model, including the purely radial LSB, the radial/temporal LSB or the axial/temporal LSB \cite{Bertolami2005,Bertolami20051}. In recent years, Casana et al. presented a Schwarzschild-like bumblebee black hole solution \cite{Casana2018}. After that, a series of other spherically symmetric solutions within the framework of the bumblebee gravity theory have been discovered, including a traversable wormhole solution \cite{Ovgun2019}, the solution with the global monopole \cite{Gullu2020}, the cosmological constant \cite{Maluf2021}, or the Einstein-Gauss-Bonnet term \cite{Ding2022}. A Kerr-like exact solution in Bumblebee gravity was obtained by Ding et al. \cite{Ding2020}, which does not seem to satisfy the corresponding field equation \cite{Kanzi2022}. However, we find that this solution can be satisfied under some certain conditions.

Linear perturbation of black holes play a important role in physics. Many astrophysical processes can be regarded as small derivations from the black hole background spacetime. For example, QNMs is considered as a good description of the late stage of the merge of binary black holes or the gravitational collapse. In order to calculate the QNMs, one need obtain the decoupled second-order differential equation in the frequency domain. For spherically symmetric spacetimes, one can always decompose the perturbed metric into spherical harmonics. And this process generally ensures that the angular part and the radial part can be automatically decoupled, while the odd-parity perturbation and the even-parity perturbation are also automatically separated \cite{ReggeWheeler1957,Zerilli1970PRD,Zerilli1970PRL,guage2022}. But this decomposition is no longer valid in rotating spacetime. One important work is Teukolsky discovered that a class of perturbation equations of Kerr spacetime can be decoupled in Newman-Penrose formalism \cite{Teukolsky1973}, the underlying reason is that the Kerr spacetime is of type-D in the Petrov classification. However, it is very difficult to decouple the perturbation equations in other rotating spacetimes, such as Kerr-Newman metric or metric for most modified gravity theories. For Proca perturbation of Kerr spacetime, the separability of perturbation equation also seems not easy \cite{Rosa2012}. In recently, Frolov et.al.
successfully solved this problem \cite{PRL2018}.

Recently, a new idea was provided by Pani et al., that is, if the slowly rotating backgrounds are close enough to a spherical symmetric spacetime, then the separability of the perturbation equations becomes possible \cite{Pani2012,Pani2012prl,Pani2013IJMPA,Pani2013prd,Pani2013prl,Tattersall2018}. This method originated from Kojima's work on the perturbations of slowly rotating neutron stars \cite{Kojima}. Under the slowly rotating limit, considering that the dimensionless rotation parameter $\tilde{a}=a/M$ of the background spacetime is sufficiently small, using the harmonic basis to expand the metric perturbation, the field equations can always reduce to a coupled ordinary differential equations. Different from the spherically symmetric case, in the limit of slowly rotating there exist a ``selection rule", i.e. the perturbations with a given parity and multipolar index $l$ are coupled to three parts \cite{Pani2012}: the perturbations with opposite parity and index $l\pm1$ at first order of $\tilde{a}$, the perturbations with the same parity and the same index $l$ up to second order of $\tilde{a}^2$, and the perturbations with the same parity and index $l\pm 2$ at the second order of $\tilde{a}$.

The aim of this manuscript is to apply the slow rotation method to Einstein-bumblebee gravity theories, and obtain the massive scalar perturbation, the Proca field perturbation and the axial gravitational perturbation for rotating Einstein-bumblebee black hole, respectively. We also calculate the QNMs of these perturbations by numerical methods, and study how the LV parameter affect the QNMs. The manuscript is organized as follows. In Sec. \ref{Sec.2}, we briefly review the Einstein-Bumblebee theory and the slow rotation technique. In Sec. \ref{Sec.3}, we derive the Schr\"{o}dinger-like equation for the massive scalar perturbation, the proca perturbation and the axial gravitational perturbation for slowly rotating Einstein-Bumblebee black hole, respectively. In Sec. \ref{QNMs}, using both the matrix method and the continued fraction method, we calculated the QNM frequencies of rotating Einstein-Bumblebee black holes. Sec. \ref{Sec7} is devoted to summary and discussion.

\section{The Theoretical Framework}\label{Sec.2}

\subsection{The Field Equations}

Here we briefly review the Einstein-Bumblebee gravity model and the black hole solutions under this gravity. This model is known as an example that extends the standard formalism of general relativity. Under a suitable potential, the bumblebee vector field $B_\mu$ acquires a nonzero vacuum expectation value and induces a spontaneous Lorentz symmetry breaking in the gravitational sector. The action for a single bumblebee field $B_\mu$ coupled to gravity can be described as \cite{Casana2018,Chen2020,wang2021}
\begin{align}\label{Action}
\mathcal{S}=&\int d^4x \sqrt{-g}\left[\frac{1}{16\pi G_N}\left( \mathcal{R}+\varrho B^aB^b \mathcal{R}_{ab} \right)\right.\non
&\left. -\frac{1}{4}B^{ab}B_{ab}-V \right]+\mathcal{L}_M,
\end{align}
$ \varrho $ is the real coupling constant which controls the non-minimal gravity interaction to bumblebee field $B_a$, and the bumblebee field strength is defined by
\begin{align}
B_{ab}=\pp_a B_b-\pp_b B_a.
\end{align}
The potential $V$ is chosen to provide a nonvanishing vacuum expectation value for bumblebee field $B_a$, and has a minimum at $B_aB^a\pm b^2$, where $b$ is a real positive constant. Hence the potential can be expressed as \cite{Casana2018}
\begin{align}
V=V(B_aB^a\pm b^2),
\end{align}
and this potential can drive a nonzero vaccum value $\langle B^a\rangle=b^a$ with $b_ab^a=\mp b^2$.

Taking the variation of \eq{Action} yields the vacuum gravitational equation and the equation of motion for the bumblebee field
\begin{align}
& \mathcal{R}_{ab}-\frac{1}{2}g_{ab}\mathcal{R}=\kappa T^{B}_{ab} \label{EinsteinEQ}, \\
& \nabla^aB_{ab}=2V'B_b-\frac{\varrho}{\kappa}B^aR_{ab}.
\end{align}
where $ \kappa=8\pi G_N $, $T^{B}_{ab}$ is the bumblebee energy momentum tensor,
\begin{align}\label{TBab}
T^{B}_{ab}=
& B_{ac}B^c\hsp_b-\frac{1}{4}g_{ab}B^{cd}B_{cd}-g_{ab}V+2B_aB_bV' \non
& \frac{\varrho}{\kappa}
\left[\frac{1}{2}g_{ab}B^{c}B^{d}R_{cd}-B_aB^cR_{cb}-B_bB^cR_{ca}\right.\non
& \left.+\frac{1}{2}\nabla_c\nabla_a\left(
B^cB_b\right)+\frac{1}{2}\nabla_c\nabla_b\left(B^cB_a\right)\right.\non
&\left.-\frac{1}{2}\nabla^2\left(B_aB_b\right)-\frac{1}{2}g_{ab}\nabla_c\nabla_d\left(B^cB^d\right)  \right],
\end{align}
and $V'$ denotes $\partial V(y)/\partial y$ at $y=B^aB_a\pm b^2$. Using the trace of Eq. \meq{EinsteinEQ}, we obtain
\begin{align}
\mathcal{R}_{ab}=&\kappa T^{B}_{ab}+2\kappa g_{ab}V-\kappa g_{ab}B^cB_cV'\non
&+\frac{\varrho}{4}g_{ab}\nabla^2\left(B^cB_c \right)+\frac{\varrho}{2}g_{ab}\nabla_c\nabla_d\left(B^cB^d \right).
\end{align}
Now further assume that the bumblebee field is fixed to be $B_a=b_a$, then the particular form of the potential is irrelevant and $V=V'=0$. Define
\begin{align}
\bar{R}_{ab}=&\mathcal{R}_{ab}-\kappa b_{ac}b^c\hsp_b+\frac{\kappa}{4}g_{ab}b^{cd}b_{cd}+\varrho b_ab^c\mathcal{R}_{cb}\non
&+\varrho b_bb^c\mathcal{R}_{ca}-\frac{\varrho}{2}g_{ab}b^cb^d\mathcal{R}_{cd}+\bar{\mathcal{B}}_{ab},
\\
\bar{\mathcal{B}}_{ab}=&\frac{\varrho}{2}\left[\nabla^2\left(b_ab_b\right)-\nabla_c\nabla_a\left(b^cb_b\right)-\nabla_c\nabla_b\left(b^cb_a\right) \right],
\end{align}
and the gravitational field Eq. \meq{TBab} is equivalent to
\begin{align}\label{fieldeq}
\bar{R}_{ab}=0.
\end{align}
In the following subsection, the satisfied of \eq{fieldeq} determines that whether the metric is the exact solution of the vacuum Einstein-Bumblebee action.

\subsection{Slowly Rotating Einstein-Bumblebee Black Hole}

Recently, using the condition $b^ab_a=constant$, and then considering the bumblebee field $b_a=(0,b_0\frac{\rho}{\sqrt{\Delta}},0,0)$, Ding et al. find the Kerr-like solution for Einstein-Bumblebee theory can be written as \cite{Ding2020}
\begin{align}\label{metricKerrlike}
ds^2=&-\left(1-\frac{2M r}{\rho^2}\right)dt^2-\frac{4M r a\sqrt{1+\ell}\sin^2\theta}{\rho^2}dtd\varphi\non
&+\frac{\rho^2}{\Delta}dr^2+\rho^2d\theta^2+\frac{\mathcal{A}\sin^2\theta}{\rho^2}d\varphi^2
\end{align}
where
\begin{align}
\rho^2=&r^2+(1+\ell)a^2\cos^2\theta, \\
\Delta=&\frac{r^2-2Mr}{1+\ell}+a^2, \\
\mathcal{A}=&\left[r^2+(1+\ell)a^2 \right]^2-\Delta(1+\ell)^2a^2\sin^2\theta .
\end{align}
and $\ell=\varrho b_0^2$ is the Lorentz-violating parameter. This metric is the solution for rotating spacetime with a radial bumblebee field. It is obvious that the metric becomes the usual Kerr metric when $\ell\rightarrow 0$ and the static Einstein-Bumblebee metric when $\tilde{a}\rightarrow 0$. Unfortunately, the metric \meq{metricKerrlike} seems not the exact-solution \cite{Kanzi2022}. Considering that $\tilde{a}$ is a sufficiently small parameter, and expanding the metric to the second order of rotation parameter $\tilde{a}$, we obtain
\begin{align}\label{metrica2}
& ds^2=g_{ab}^{(2)}dx^adx^b \non
=& -\left(1-\frac{2M}{r}+\frac{2a^2M(1+\ell)\cos^2\theta}{ r^3}\right)dt^2 \non
& -\frac{4aM\sqrt{1+\ell}\sin^2\theta}{r}dtd\varphi+\frac{(1+\ell)\rho^2}{\tilde{\Delta}}dr^2 \non
& +\left[r^2+a^2(1+\ell)\cos^2\theta \right]d\theta^2 \non
& +\sin^2\theta\left[r^2+a^2\left(1+\ell\right)\left(1+\frac{2M}{r}\sin^2\theta\right) \right]d \varphi^2,
\end{align}
Note that $\tilde{\Delta}$ is equivalent to $\Delta$ in the second order of $\tilde{a}$, and which can determine the singular points $r_{+}$ and $ r_{-}$ as the roots of $\tilde{\Delta}$
\begin{align}
\tilde{\Delta}=(r-r_{+})(r-r_{-}).
\end{align}
where
\begin{align}\label{rpm}
r_{+}=2M-(1+\ell)\frac{a^2}{2M},~~~~r_{-}=(1+\ell)\frac{a^2}{2M}.
\end{align}
We find that this metric yields some components of the field equation do not satisfied. However, we find the non vanishing terms for field equations are all proportional to $\ell^2\tilde{a}^2$. Under certain conditions, this metric can be considered as an approximate solution of the Einstein-Bumblebee Eq. \meq{fieldeq}. The detailed discussions are presented in Appendix \ref{AppendixA}.

\section{Perturbations of Slowly Rotation Einstein-Bumblebee Black Hole}\label{Sec.3}
In this section, using the slow rotation approximation, we analysis the perturbations of Einstein-Bumblebee black hole. For massive scalar field perturbation, our analysis up to the second order of $\tilde{a}$. And for massive vector field perturbation or odd-parity gravitational perturbation, our analysis only up to the first order of $\tilde{a}$. Here we briefly introduce the notations of the spherical harmonic basis \cite{Thorne1980}. First we define two unnormalized and orthogonal co-vectors $v$ and $n$ as
\begin{equation}
v_a=(-1,0,0,0), \quad\quad n_a=(0,1,0,0),
\end{equation}
the projection operator onto the sphere surface
\begin{equation}
\Omega_{ab}=r^2 \text{diag}(0,0,1,\sin^2\theta),
\end{equation}
and the spatial Levi-Civita tensor, $\epsilon_{abc}\equiv v^d\epsilon_{dabc}$, where $\epsilon_{tr\theta\phi}=r^2\sin\theta$. using the scalar spherical harmonics $Y^{lm}=Y^{lm}(\theta,\varphi)$, the pure-spin vector harmonics are given by
\begin{align}\label{vectorbasis}
Y_a^{E,lm}&= r\grad_a Y^{lm}, \non
Y_a^{B,lm}&= r\epsilon_{ab}\hsp^c n^b\grad_c Y^{lm}, \non
Y_a^{R,lm}&= n_aY^{lm},
\end{align}
and the pure-spin tensor harmonics are given by
\begin{align}\label{tensorbasis}
T_{ab}^{T0,lm}&= \Omega_{ab}Y^{lm}, \non
T_{ab}^{L0,lm}&= n_an_b Y^{lm}, \non
T_{ab}^{E1,lm}&= rn_{(a}\grad_{b)}Y^{lm}, \non
T_{ab}^{B1,lm}&= rn_{(a}\epsilon_{b)c}\hsp^d n^c\grad_d Y^{lm}, \non
T_{ab}^{E2,lm}&= r^2\left(\Omega_a\hsp^c\Omega_b\hsp^d-\frac{1}{2}\Omega_{ab}\Omega^{cd}\right)\grad_c\grad_dY^{lm}, \non
T_{ab}^{B2,lm}&= r^2\Omega_{(a}\hsp^c\epsilon_{b)e}\hsp^d n^e\grad_c\grad_d Y^{lm}.
\end{align}
These spherical harmonic basis will be used to decompose vector or tensor perturbations. For axially symmetric background, the perturbation with different values of $m$ are decoupled, therefore we ignore the index $m$ in the following discussions. And the linearized perturbed equations always imply a sum over $(l,m)$. Some identities will be used, which are given by \cite{Kojima}
\begin{align}
\cos\theta Y^{l}=&\mathcal{Q}_{l+1}Y^{l+1}+\mathcal{Q}_{l}Y^{l-1}, \label{identities1}\\
\sin\theta \pp_\theta Y^{l}=&\mathcal{Q}_{l+1}lY^{l+1}-\mathcal{Q}_{l}(l+1)Y^{l-1}, \label{identities2}\\
\cos^2\theta Y^{l}=&\left(\mathcal{Q}^2_{l+1}+\mathcal{Q}^2_{l}\right)Y^{l} \non
& +\mathcal{Q}_{l+1}\mathcal{Q}_{l+2}Y^{l+2} +\mathcal{Q}_{l}\mathcal{Q}_{l-1}Y^{l-2}, \label{identities3}\\
\cos\theta \sin\theta \pp_\theta Y^{l}=&\left[l\mathcal{Q}^2_{l+1}-(l+1)\mathcal{Q}^2_{l}\right]Y^{l}\non
&+\mathcal{Q}_{l+1}\mathcal{Q}_{l+2}lY^{l+2}\non
&-\mathcal{Q}_{l}\mathcal{Q}_{l-1}(l+1)Y^{l-2} \label{identities4},
\end{align}
where
\begin{align}
\mathcal{Q}_{l}=\sqrt{\frac{l^2-m^2}{4l^2-1}}.
\end{align}

\subsection{Massive scalar perturbation}\label{secscalar}
Considering that the scalar field coupling to the bumblebee field is neglected, then the massive Klein-Gordon equation reads
\begin{align}\label{KGequation}
&\frac{1}{\sqrt{-g}}\pp_a\left(\sqrt{-g}g^{ab}\pp_b\phi\right)=\mu^2\phi,
\end{align}
where $ m_s=\mu\hbar $ represents the scalar field's mass. We decompose the field in spherical harmonics:
\begin{align}
\phi=\sum_{lm}\frac{\Psi_{l}(r)}{\sqrt{r^2+a^2}}e^{-i\omega t}Y^{lm}(\theta,\varphi).
\end{align}
Since in axially symmetric background, perturbations with different values of $m$ are decoupled, the index of $m$ can be neglected. Substituting the metric \meq{metrica2} into Eq. \meq{KGequation} and expanding the equation up to the second order of $\tilde{a}$, we obtain
\begin{align}\label{AlDl}
A_lY^{l}+a^2 D_l\cos^2\theta\cdot Y^{l}=0,
\end{align}
We find that the parameters $A_l$ and $D_l$ are constructed by $\Psi_l$ and its derivatives, and present their specific expression in Appendix \ref{AppendixB}. It shows that in \eq{AlDl} the radial and the angular sections are still coupled together, hence one needs to decouple the angular section to obtain a purely radial equation.

Expanding \eq{AlDl} up to the first order of $\tilde{a}$, one can obtain
\begin{align}\label{scalara1}
\frac{d^2}{dx^2}\Psi_l^{(1)}+V^{(1)}_l\Psi_l^{(1)}=0,
\end{align}
where $dr/dx=\mathcal{F}=1-2M/r $, and the effective potential is
\begin{align}
V^{(1)}_l=&(1+\ell)\left(\omega^2-\sqrt{1+\ell}\frac{4aMm\omega}{r^3}\right)\non
&-\mathcal{F}\left[\frac{2M}{r^3}+(1+\ell)\left(\frac{l(l+1)}{r^2}+\mu^2\right)\right],
\end{align}
to represent the potential expanded up to the first order of $\tilde{a}$. When the Lorentz-violating parameter $\ell$ vanishes, the effective potential will be consistent with Eq. \meq{identities2} in Ref. \cite{Pani2012}.

Now we separate the angular part of \eq{AlDl} up to the second order of $\tilde{a}$. Using the identity \eq{identities3} and taking the inner product of the Eq. \meq{AlDl} with the conjugate of scalar spherical harmonics on the sphere, we can get
\begin{align}\label{fieldEQ}
A_l+&a^2\left[\left(\mathcal{Q}^2_{l+1}+\mathcal{Q}^2_l\right)D_l+\mathcal{Q}_{l-1}\mathcal{Q}_lD_{l-2}\right.\non
&\left.+\mathcal{Q}_{l+2}\mathcal{Q}_{l+1}D_{l+2}\right]=0,
\end{align}
the parameters $ D_{l\pm2} $ is made up of $ \Psi_{l\pm2} $ and its derivatives, which are written as
\begin{align}
\label{Dlpm2}
D_{l\pm 2}=\frac{1}{r^2(2M-r)}\left(\frac{d^2}{d x^2}\Psi_{l\pm 2}+W_{l\pm 2}\Psi_{l\pm 2} \right).
\end{align}
Note that the explicit expression of $ W_{l\pm 2} $ is unnecessary. The reason is that all $D_{l\pm 2}$ terms is proportional to $\tilde{a}^2$, one should only consider $D_{l\pm2}$ at the zeroth order of $\tilde{a}$. This implies that the functions $\Psi_{l\pm2}$ in \eq{Dlpm2} can be only considered as the solutions of
\begin{align}\label{scalara0}
\frac{d^2}{dx^2}\Psi_{l\pm 2}^{(0)}+V^{(0)}_{l\pm 2}\Psi_{l\pm 2}^{(0)}=0.
\end{align}
where $V^{(0)}=V^{(1)}|_{a=0}$. And the expression of $D_{l\pm2}$ at the zeroth order can be obtained,
\begin{align}
D_{l\pm 2}^{(0)}=\frac{(1+\ell)(\omega^2-\mu^2)}{r^3}\Psi_{l\pm 2}^{(0)},
\end{align}
Then, using the explicit form of the coefficients $A_{l\pm2}$ and $D_{l\pm2}$, the field equation\meq{fieldEQ} has the form as
\begin{align}\label{scalara2}
&\frac{d^2}{dr_*^2}\Psi_l^{(2)}+V_l^{(2)}\Psi_l^{(2)}=\frac{a^2\mathcal{F}(1+\ell)^2(\mu^2-\omega^2)}{r^2}\non
&\times\left[\mathcal{Q}_{l+1}\mathcal{Q}_{l+2}\Psi_{l+2}^{(0)}+\mathcal{Q}_{l-1}\mathcal{Q}_{l}\Psi_{l-2}^{(0)} \right],
\end{align}
where $ dr/dr_*=f $,
\begin{align}
f=\frac{(1+\mathcal{\ell})\tilde{\Delta}}{r^2+a^2},
\end{align}
and the potential is given by
\begin{align}
V_l^{(2)}=&V^{(1)}_l+\frac{a^2}{r^6}\Big[-24M^2+2Mr\big(6-2l(l+1)(1+\ell) \non
& -3\ell-2r^2(1+\ell) \mu^2+r^2(1+\ell)^2\omega^2  \big) \non
& +r^2\big(l-1+2\ell-l\ell^2+m^2(1+\ell)^2-l^2(\ell^2-1) \non
& +r^2\mu^2-r^2\left(\omega^2+\ell^2\mu^2-\ell^2\omega^2 \right) \big)\non
& -r^4\mathcal{F}(1+\ell)^2(\mu^2-\omega^2)\left(\mathcal{Q}_{l}^2+\mathcal{Q}_{l+1}^2\right)
\Big].
\end{align}
This potential coincides with Pani's result when $\ell\rightarrow 0$. To obtain the single variable equation for $\Psi_l$, we need define
\begin{align}
Z_l=\Psi_l^{(2)}+a^2c_l \Psi_{l-2}^{(2)}-a^2c_{l+2}\Psi_{l+2}^{(2)},
\end{align}
where $ c_l $ reads as
\begin{align}
c_l=\frac{(1+\ell)}{2(2l-1)}\left(\mu^2-\omega^2\right)\mathcal{Q}_{l-1}\mathcal{Q}_{l}.
\end{align}
Then, the wave-like equation for scalar field perturbation is given by
\begin{align}\label{mastereqss}
\frac{d^2}{dr_*^2}Z_l+V_l^{(2)}Z_l=0,
\end{align}
and which can be analyzed by standard methods.

\subsection{Massive Vector Perturbation}\label{Sec4}

The field equations of a massive vector field, or known as Proca field, is
\begin{align}\label{deltamaxwell}
\Pi^b=\nabla_aF^{ab}-\mu^2A^b=0,
\end{align}
where $A^a$ is the vector potential, $\mu=m_v/\hbar$ is the mass of the vector field and $F_{ab}=2\nabla_{[a}A_{b]}$. We also consider this vector field satisfied the Lorentz condition $\nabla_aA^a=0 $. With the help of the vector spherical harmonic basis \meq{vectorbasis}, we expand the perturbed vector $\delta A_a$ as,
\begin{align}
\delta A_a=&-\frac{\mathrm{P}_l}{r}v_aY^{lm}+\frac{\mathrm{R}_l}{rf}Y^{R,lm}_a\non
&+\frac{\mathrm{S}_l}{r\Lambda}Y^{E,lm}_a
-\frac{\sqrt{1+\ell} \mathrm{Q}_l}{r\Lambda}Y^{B,lm}_a,
\end{align}
where $\Lambda=l(l+1)$, and $\mathrm{P}_l$, $\mathrm{R}_l$, $\mathrm{S}_l$, $\mathrm{Q}_l$ are all scalar functions of coordinate $(t,r)$. The functions $\mathrm{P}_l$, $\mathrm{R}_l$, $\mathrm{S}_l$ belong to the polar sector and the function $\mathrm{Q}_l$ belongs to the axial sector. Moreover, assume an harmonic time-dependence in time, for instance,
\begin{align}\label{scalara22}
\mathrm{P}_l(t,r)=e^{-i \omega t}\mathrm{P}_l(r).
\end{align}
The perturbation of the four components of Eq.\meq{deltamaxwell} and the Lorentz condition can be naturally separated into three groups:
\begin{align}\label{group1}
\delta\Pi_{I}=&\left(A^{(I)}_l+\tilde{A}^{(I)}_l\cos\theta+D^{(I)}_l\cos^2\theta \right)Y^l\non
&+\left(B^{(I)}_l+\tilde{B}^{(I)}_l \cos^2\theta  \right)\sin\theta \frac{\pp}{\pp \theta}Y^l=0,
\\
\label{group2}
\delta \Pi_\theta =&\left(\alpha_l+\rho_l\sin^2\theta  \right)\frac{\pp}{\pp\theta}Y^l-i\hsp m \beta_l
\frac{Y^l}{\sin\theta}\non
&+\left(\eta_l+\sigma_l\cos\theta\right)\sin\theta Y^l=0,
\\
\label{group3}
\frac{\delta \Pi_\varphi}{\sin\theta}=&\left(\beta_l+\gamma_l\sin^2\theta \right)\frac{\pp}{\pp \theta}Y^l+i\hsp m \alpha_l\frac{Y^l}{\sin\theta}\non
&+\left(\zeta_l+\lambda_l\cos\theta \right)\sin\theta Y^l=0.
\end{align}
The index $I=\{t,r,L\}$ represents the $t$ component, the $r$ component of \eq{group1}, and the expansion of the Lorentz condition $\nabla_aA^a=0$, respectively. The explicit expression of the coefficients in \eqs{group1} to \meq{group3} are presented in Appendix \ref{AppendixC}.

For Eq. \meq{group1}, the angular part can be separated by taking the inner product to $Y_{lm}^*$. But for \eqs{group2} and \meq{group3}, in order to completely separate the angular part, we need construct a new vector $\tilde{\Pi}_a$ as
\begin{align}
\tilde{\Pi}_a\equiv\left(0,0,\delta\Pi_\theta,\delta\Pi_\varphi\right).
\end{align}
Taking the inner product with $Y^{a*}_{E,l'm'}$ and $Y^{a*}_{B,l'm'}$, respectively,
\begin{align}
\int \tilde{\Pi}_a Y^{a*}_{E,l'm'}d\Omega, \\
\int \tilde{\Pi}_a Y^{a*}_{B,l'm'}d\Omega,
\end{align}
and also using the identities \eqs{identities1}-\meq{identities4}, then the radial equations can be obtained. We list these radial equations in Appendix \ref{AppendixC}, see \eqs{MEQtrL}-\meq{MEQvarphi}.

First we consider the static Proca field equations. When $\tilde{a}\rightarrow0$, \eqs{MEQtrL}-\meq{MEQvarphi} yield
\begin{align}\label{D2R}
&\hat{\mathcal{D}}_2 \mathrm{R}_l-\frac{2\mathcal{F}}{r^2}\left(1-\frac{3M}{r}\right)
\left[\mathrm{R}_l-(1+\ell)\mathrm{S}_l \right]=0,\\ \label{D2S}
&\hat{\mathcal{D}}_2 \mathrm{S}_l+\frac{2 \Lambda \mathcal{F}}{r^2}\mathrm{R}_l=0,\\ \label{D2Q}
&\hat{\mathcal{D}}_2 \mathrm{Q}_l=0,
\end{align}
where the operator $\hat{\mathcal{D}}_2$ is introduced as
\begin{align}\label{hatD2}
\hat{\mathcal{D}}_2=\frac{d^2}{dx^2}+(1+\ell)\left[\omega^2-\mathcal{F}\left(\frac{\Lambda}{r^2}+\mu^2\right)\right].
\end{align}
It shows that in this case, two sectors are naturally decoupled. Using \eq{D2S} to eliminate $\mathrm{R}_l$ in \eq{D2R}, one can obtain
\begin{align}\label{D2D2S}
\hat{\mathcal{D}}_2\left[\frac{r^2}{\mathcal{F}}\hat{\mathcal{D}}_2\mathrm{S}_l \right]
-\left(1-\frac{3M}{r}\right)\left[2\hat{\mathcal{D}}_2 +(1+\ell)\frac{4\Lambda\mathcal{F}}{r^2} \right]\mathrm{S}_l=0,
\end{align}
which is a fourth order differential equation. Note that we can classify the eigenvectors of the system according to the three degrees of freedom of the vector $A_a$, i.e., the three polarizations, which include one scalar type polarization and two vector type polarizations. And the electric mode of the vector potential $A_a$ has one scalar type polarization and one vector type polarization. The reason of the above equation as a fourth order differential equation is that, $\mathrm{S}_l$ contains both scalar type polarization and vector type polarization.

Now we consider the situation of expanding the field equations to the first order of $\tilde{a}$. The polar sector of the field equations gives
\begin{align}\label{D2Ra1}
&\hat{\mathcal{D}}_2 \mathrm{R}_l-\frac{2\mathcal{F}}{r^2}\left(1-\frac{3M}{r}\right)\bigg[\mathrm{R}_l-(1+\ell)\mathrm{S}_l \bigg]=\mathcal{S}^{\text{po}}_1,
\\ \label{D2Sa1}
&\hat{\mathcal{D}}_2\mathrm{S}_l+\frac{2\Lambda \mathcal{F}}{r^2}\mathrm{R}_l=\mathcal{S}^{\text{po}}_2,
\end{align}
where $ \mathcal{S}^\text{po}_1 $ and $ \mathcal{S}^\text{po}_2 $ are the source terms, and the explicit form of which are presented in Appendix \ref{AppendixC}. While the axial sector of the field equations gives
\begin{align}\label{D2Qa1}
\hat{\mathcal{D}}_2 \mathrm{Q}_l-(1+\ell)^{\frac{3}{2}}\frac{4a\hsp Mm\omega}{r^3}\mathrm{Q}_l=\mathcal{S}^{\text{ax}}.
\end{align}
where $\mathcal{S}^\text{ax}$ is the source term and given in Appendix \ref{AppendixC}. It shows that the expressions of  $\mathcal{S}^\text{po}_1$, $\mathcal{S}^\text{po}_2$ and $\mathcal{S}^\text{ax}$ are proportional to $\tilde{a}$, therefore it is only necessary to consider the components of $\mathcal{S}$, such as $\mathrm{Q}_l$ or $\mathrm{R}_l$, to the zeroth order of $\tilde{a}$.

Note that for the first order rotation approximation, the massive scalar perturbation and the massive vector perturbation can be uniformly written as
\begin{align}\label{meqspin}
\hat{\mathcal{D}}_2 \Psi_l-\frac{2M}{r^3}\left[(1-s^2)\mathcal{F}+2a\hsp m(1+\ell)^\frac{3}{2}\omega  \right]\Psi_l=0.
\end{align}
where $s=0$ for scalar perturbations and $s=\pm 1$ for vector perturbations with axial parity. However, from the nextsubsection, it shows that in Einstein-Bumblebee theory, the gravitational perturbation with axial parity can not be uniformly written in the form of the above equation.

\subsection{Gravitational Perturbation}\label{Grafield}

We consider that the metric $g_{ab}$ can be decomposed into the background $g^{(0)}_{ab}$ and the perturbation $h_{ab}$
\begin{align}
g_{ab}=g^{(0)}_{ab}+h_{ab}.
\end{align}
Using the scalar spherical harmonic, the pure-spin vector harmonics \meq{vectorbasis} and the tensor harmonics \meq{tensorbasis}, one can obtain ten orthogonal spherical harmonic basis \cite{Thorne1980}. And the metric perturbation can be decomposed as \cite{Thompson2016}
\begin{align}
h_{ab}^{lm} =& \mathrm{A}_l~v_av_bY^{lm}+2\mathrm{B}_l~v_{(a}Y_{b)}^{E,lm}+2\mathrm{C}_l~v_{(a}Y_{b)}^{B,lm}\non
&+2\mathrm{D}_l~v_{(a}Y_{b)}^{R,lm}+\mathrm{E}_l~T^{T0,lm}_{ab}  +\mathrm{F}_l~T_{ab}^{E2,lm}\non
&+\mathrm{G}_l~T^{B2,lm}_{ab}+2\mathrm{H}_l~T^{E1,lm}_{ab}+2\mathrm{J}_l~T^{B1,lm}_{ab}\non
&+\mathrm{K}_l~T^{L0,lm}_{ab} .
\end{align}
where the coefficients $\mathrm{A}_l$ to $\mathrm{K}_l$ are scalar functions. Adopting the well known Regge-Wheeler gauge \cite{ReggeWheeler1957,Thompson2016,Jing2022,Gerlach1979,Gerlach1980,Poisson2005,Poisson2006,Poisson2018}, we set
\begin{equation}
\mathrm{B}_l=\mathrm{F}_l=\mathrm{H}_l=\mathrm{G}_l=0,
\end{equation}
After separating the angular components of \eq{fieldeq}, ten pure radial equations can be obtained, which are naturally separated into axial parity and polar parity. The two independent components for the axial parity equation are given by

\begin{align}\label{R13}
\bar{R}_{t\varphi}:~~
& \frac{\Lambda\mathcal{F}}{2}
\left(2\mathrm{C}_l'+r\mathrm{C}_l''-3i\omega\mathrm{J}_l
-ir\omega\mathrm{J}_l'\right)+\frac{aMm\sqrt{1+\ell}}{r^2}\non
&\times\left[\frac{2-\Lambda}{\mathcal{F}}\omega\mathrm{C}_l
+i\left(\frac{12M}{r^2}\mathrm{J}_l-\frac{4+\Lambda}{r}\mathrm{J}_l+2\mathcal{F}\mathrm{J}_l'\right)  \right]\non
&+\frac{\Lambda(4M-r\Lambda)}{2r^2}\mathrm{C}_l=0,
\end{align}
\begin{align}
\label{R23}
\bar{R}_{r\varphi}:~~
&\frac{(1+\ell)\Lambda}{2\mathcal{F}}\left(\frac{2(2-\Lambda)M}{r^2}+\frac{\Lambda-2-r^2\omega^2}{r} \right)\mathrm{J}_l\non
&+\frac{i\omega(1+\ell)\Lambda}{2\mathcal{F}}\left(\mathrm{C}_l-r\mathrm{C}_l' \right)+\frac{aMm(1+\ell)^{\frac{3}{2}}}{\mathcal{F}}\non
&\times\left[i\frac{6-\Lambda}{r^3}\mathrm{C}_l+\frac{\Lambda}{r^2}\left(2\omega\mathrm{J}_l+i\mathrm{C}_l'  \right) \right]=0.
\end{align}
Together with \eqs{R23} and \meq{R13}, define $\mathcal{U}_l$ as
\begin{align}
\mathrm{J}_l=\frac{1}{\mathcal{F}}\left(1-2\sqrt{1+\ell}\frac{aMm}{r^3\omega}  \right)\mathcal{U}_l,
\end{align}
then the modified master equation describing the axial gravitational perturbation up to the first order of $\tilde{a}$ can be written as
\begin{align}\label{graveq}
\frac{d^2}{d x^2}\mathcal{U}_l+\left[\omega^2-\mathcal{V}_l\right]\mathcal{U}_l=0,
\end{align}
where
\begin{align}\label{gravV}
\mathcal{V}_l=\mathcal{F}\left(\frac{\Lambda}{r^2}-\frac{6M}{r^3} \right)+&\sqrt{1+\ell}\frac{4aMm}{r^3}\non
&\times\left[\omega+6\mathcal{F}\left(\frac{3r-7M}{r^3\Lambda\omega}  \right)  \right].
\end{align}
From the expression of the above potential, we find that the Lorentz-violating parameter coupled with rotation parameter $\tilde{a}$, which implies that the axial parity perturbation equation for the static black hole solution in Einstein-Bumblebee theory is the same as Schwarzschild black hole.

\section{The Eigenvalue Problem For Quasinormal Modes}\label{QNMs}
\subsection{Boundary conditions}

In this manuscript, we are concerned about how the Lorentz-violating parameter $\ell$ affects the QNMs. Hence we only investigate the QNMs at the massless limit. At the first order of $\tilde{a}$, the relation \meq{rpm} shows that there is only one horizon at $r_{+}=2M$. For scalar field or vector field, the generic wave function $\Psi_l$ has the asymptotic behavior as
\begin{equation}\label{solvesv}
\Psi_l\sim
\begin{cases}
\left(r-2M\right)^{-i \tilde{\ell}\hsp\sigma_+} &\text{for}~~r\rightarrow~2M ,\\
e^{i\tilde{\ell}\hsp\omega x} &\text{for}~~r\rightarrow~+\infty ,
\end{cases}
\end{equation}
where $x$ is the tortoise coordinate given in Eq. \meq{scalara1}, $\tilde{\ell}$ and $ \sigma_{+}$ are defined as
\begin{align}
\tilde{\ell}=\sqrt{1+\ell}, \quad \sigma_{+}=2M\omega-\tilde{\ell}\frac{m a}{2M}.
\end{align}
But for the gravitational field Eq. \meq{gravV}, the generic wave function $\mathcal{U}_l$ has the asymptotic behavior as
\begin{equation}
\mathcal{U}_l\sim
\begin{cases}
\left(r-2M\right)^{-i \sigma_+} &\text{for}~~r\rightarrow~2M ,\\
e^{i\omega x} &\text{for}~~r\rightarrow~+\infty .
\end{cases}
\end{equation}
Using these asymptotic solutions, in order to apply the numerical method, we can impose $\Psi_l$ and $\mathcal{U}_l$ satisfied the relation as
\begin{align}\label{psisv}
\Psi_l= & e^{i\tilde{\ell}\omega r}r^{i\tilde{\ell}\left(2M \omega+\sigma_+\right)}\left(r-2M\right)^{-i\tilde{\ell}\hsp\sigma_+}\psi_l,
\\\label{psig}
\mathcal{U}_l= &e^{i\omega r}r^{i\left(2M \omega+\sigma_+\right)}\left(r-2M\right)^{-i\sigma_+}\psi_l.
\end{align}

If we consider the case that up to the second order of rotation parameter $\tilde{a}$, two horizons of the rotation black hole are determined by the \eq{rpm}.
At this case the asymptotic behavior of $Z_l$ can be written as
\begin{equation}\label{solves2}
Z_l\sim
\begin{cases}
\left(r-r_+\right)^{-i \tilde{\ell}\hsp\Omega} &\text{for}~~r\rightarrow~r_+ ,\\
e^{i\tilde{\ell}\hsp\omega r_*} &\text{for}~~r\rightarrow~+\infty ,
\end{cases}
\end{equation}
where
\begin{align}
r_{+}=2M-(1+\ell)\frac{a^2}{2M}, \quad \Omega=\left(4M-r_+\right)\omega-\tilde{\ell}\hsp \frac{ma}{2M}.
\end{align}
According to this asymptotic behavior, $Z_l$ can be assumed to be written as
\begin{align}\label{psia2}
Z_l= & e^{i\tilde{\ell}\omega r}r^{i\tilde{\ell}\left(2M \omega+\Omega\right)}\left(r-r_+\right)^{-i\tilde{\ell}\hsp\Omega}\psi_l.
\end{align}

In the following sections, we will analyze QNMs by using the matrix method and the continued fraction method, respectively.

\subsection{Matrix method for quasinormal modes}

To compute the QNMs, here we briefly describe the matrix method presented by Lin et.al. \cite{Lin20171,Lin2017,Lin2019,Lei2021}. By taking into account the boundary conditions mentioned in the previous subsection, we perform a coordinate transformation
\begin{align}\label{transyr}
y=1-\frac{r_+}{r},
\end{align}	
so the region of QNMs calculation becomes $ y\in[0,1]$. Considering the boundary condition, assuming the wave function can be reconstructed as
\begin{align}\label{transchipsi}
\chi(y)=y(1-y)\psi_l(y),
\end{align}
then the boundary condition at the event horizon and the spatial infinity becomes
\begin{align}\label{chi01}
\chi(0)=\chi(1)=0.
\end{align}
This boundary condition ensures that the resulting matrix equation is homogenous, as will be seen below. It can be proved that all perturbation field equations can be rewritten as
\begin{align}\label{qicieq}
\mathcal{C}_2(y,\omega)\chi''(y)+\mathcal{C}_1(y,\omega)\chi'(y)+\mathcal{C}_0(y,\omega)\chi(y)=0,
\end{align}
where the functions $\mathcal{C}(y,\omega)$ can be derived by substituting the behavior of the wave function into the corresponding field equations. For instance, together with \eqs{transyr} and \meq{transchipsi}, one can substitute \eq{psia2} into \meq{mastereqss} and obtain the equation in the form of \eq{qicieq}. Note that all $\mathcal{C}_j(j=0,1,2)$ are linear functions of $\omega$, so $\mathcal{C}_j$ can be decomposed as $\mathcal{C}_j(y,\omega)=\mathcal{C}_{j,0}(y)+\omega\hsp \mathcal{C}_{j,1}(y)$.

Using the matrix method to discretize \eq{qicieq}, we introduce equally spaced grid points into the internal $[0,1]$.
By expanding the function $ \chi(y) $ around each grid point using the Taylor series, the corresponding differential matrices can be constructed.
Thus, Eq. \meq{qicieq} is rewritten as an algebraic equation in matrix form
\begin{align}
\left(\mathcal{M}_0+\omega\mathcal{M}_1\right)\chi(y)=0,	
\end{align}
where $ \mathcal{M}_0 $ and $ \mathcal{M}_1 $ are matrices consisting of the functions $\mathcal{C}_j$ and the corresponding differential matrices. Calculating these matrices \cite{Lin20171,Lin2017}, and then the solve of QNMs becomes a simple algebraic solution problem.

\subsection{Continued fraction method}\label{Sec.5}
Since the groundbreaking work by Leaver \cite{Leaver1985}, the continued fraction method is an accurate method in determining the QNMs. In this method, the eigenfunction can be expressed as a series whose coefficients adhere to a finite term recurrence relation. In order to obtain a more concise recurrence relation, in this subsection we set $M=1/2$.

A solution to the perturbation equation that expands at the event horizon can be written in the following form:
\begin{align}
\psi_l=\sum_{n=0}^{\infty}d_n\left(\frac{r-r_+}{r-r_-} \right)^n.
\end{align}
For the Eq. \meq{meqspin} controlling the scalar and electromagnetic fields, the expansion coefficients are defined by a three-term recursion relations
\begin{align}
[1+i(1+\ell)&m\tilde{a}-2i\tilde{\ell}\omega]d_1+\big[s^2-1-(1+\ell)(\Lambda-8\omega^2)\non &+4i\tilde{\ell}\omega-(1+\ell)m\tilde{a}(i+3\tilde{\ell}\omega)\big]d_0=0,
\end{align}
\begin{align}\label{3oderditui}
d_{n+1}\alpha_n+d_n\beta_n+d_{n-1}\gamma_n=0,&&n=1,2,3\cdots
\end{align}
The recurrence coefficients $ \alpha_n $, $\beta_n  $, and $ \gamma_n $ are simple functions consisting of $ n $ and other differential equation parameters, the explicit forms are as follows
\begin{align}
\alpha_n=&4(1+n)(1+n+i(1+\ell)m\tilde{a}-2i\tilde{\ell}\omega),\\
\beta_n=&4(s^2-1-2n^2)-8n[1+i(1+\ell)m\tilde{a}-4i\tilde{\ell}\omega]\non
&-4\tilde{\ell}[\tilde{\ell}\Lambda+\tilde{\ell}m\tilde{a}(i+3\tilde{\ell}\omega)]+16\tilde{\ell}\omega(i+2\tilde{\ell}\omega),
\\
\gamma_n=&4(n^2-s^2)+4in[(1+\ell)m\tilde{a}-4\tilde{\ell}\omega]\non
&+8(1+\ell)(\tilde{\ell}m\tilde{a}-2\omega)\omega.
\end{align}
The quasi-normal frequencies can be obtained by solving the algebraic \eq{3oderditui} at sufficiently large $n$ for any initial value $d_0$. Without loss of generality, we set $d_0=1$.

For the case of gravitational perturbations \eq{gravV}, the recursion relations appear to be more complicated. The explicit form of these relations are given by
\begin{align}\label{grav0-4}
d_1=&\tilde{\mathcal{C}}_{1,0}~d_0,\non
d_2=&\tilde{\mathcal{C}}_{2,0}~d_0+\tilde{\mathcal{C}}_{2,1}~d_1,\non
d_3=&\tilde{\mathcal{C}}_{3,0}~d_0+\tilde{\mathcal{C}}_{3,1}~d_1+\tilde{\mathcal{C}}_{3,2}~d_2,\\
d_4=&\tilde{\mathcal{C}}_{4,0}~d_0+\tilde{\mathcal{C}}_{4,1}~d_1+\tilde{\mathcal{C}}_{4,2}~d_2+\tilde{\mathcal{C}}_{4,3}~d_3,\nonumber
\end{align}
\begin{align}\label{6oder}
&d_{n+1}\alpha_n+d_{n}\beta_n+d_{n-1}\gamma_n \non
&+d_{n-2}\sigma_n+d_{n-3}\tau_n+d_{n-4}\delta_n=0,&&n=4,5,6\cdots
\end{align}
where $\tilde{\mathcal{C}}_{i,j}$, as well as $ \alpha_n $, $ \beta_n $, $ \gamma_n $, $ \sigma_n $, $ \tau_n $ and $ \delta_n $, are functions consisting of $n$ and other parameters. The explicit form are presented in Appendix \ref{AppendixD}.

\subsection{Numerical results}

Using the matrix method and the continued fraction method, we numerically calculated QNMs and show the result in this subsection. In matrix method, we set $N=18$, to ensure that the relative error becomes smaller than $10^{-5}$. In continuous fraction method, we computed the 60th order for scalar or electromagnetic perturbations and 16th order for gravitational perturbation. The reason is that the latter equation is more complex and consumes too much computing resources. However, the numerical results shows that the difference between two methods is smaller than $10^{-4}$. In the calculating of QNMs, we set $M=1$. In this paper we only provide data for the $l=m=2$ modes, which expected to be the most astrophysically relevant. The detailed calculation data can be found in Appendix \ref{AppendixE}.

\subsubsection{Static of Bumblebee modes}

\begin{figure}[H]
\centering
\includegraphics[width=0.9\linewidth]{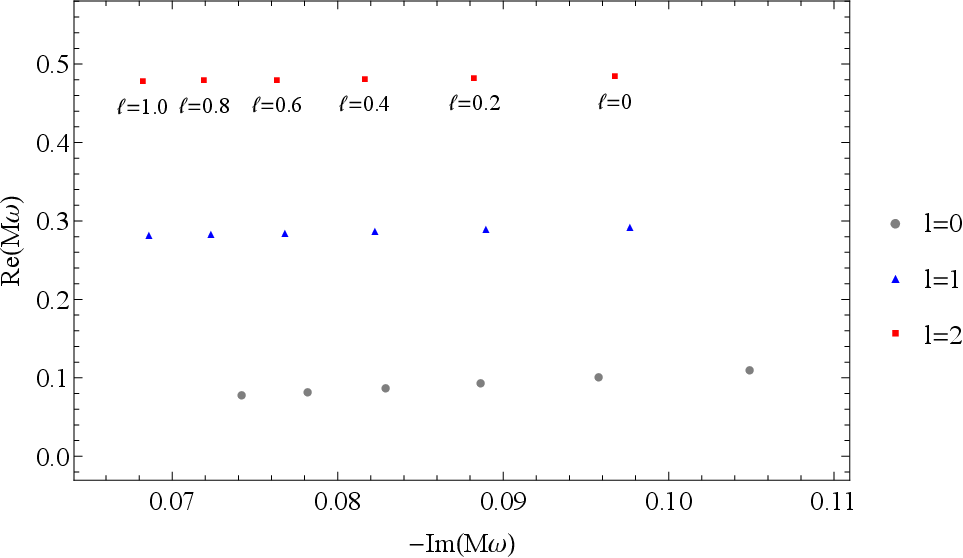}
\caption{Complex massless scalar frequencies for the $n=0$ mode for varying values of $\ell$.}
\label{figs}
\end{figure}

\begin{figure}[H]
\centering
\includegraphics[width=0.9\linewidth]{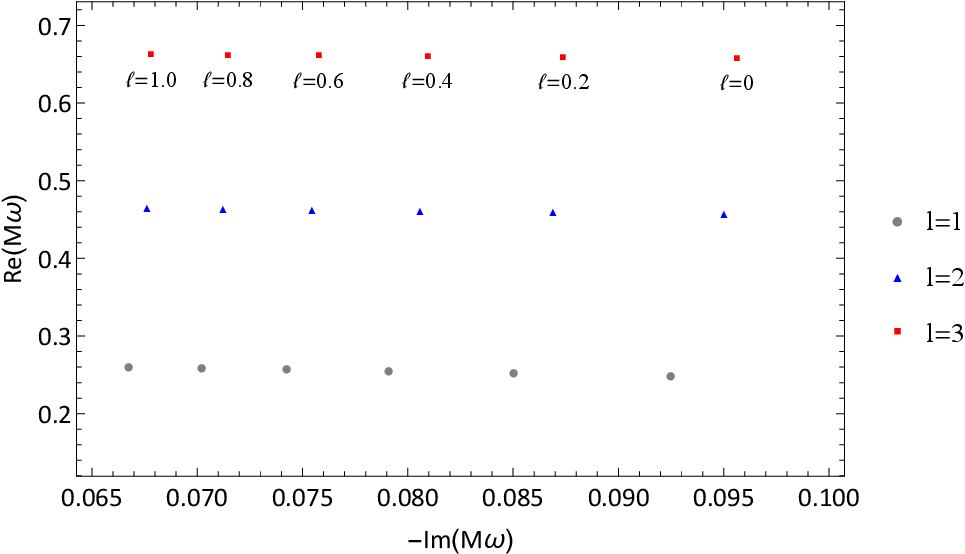}
\caption{Complex massless electromagnetic frequencies for the $n=0$ mode for varying values of $\ell$.}
\label{figv}
\end{figure}

First, we consider the static black hole in the Einstein-Bumblebee theory. When $\tilde{a}\rightarrow 0$, the metric \meq{metricKerrlike} becomes the static Schwarzschild-like black hole solution given by Ref. \cite{Casana2018}. Note that this static Einstein-Bumblebee solution is an exactly solution without any approximation. From the determinant of the metric \meq{metricKerrlike}, in order to maintain the Lorentz signature, the parameter $\ell$ should satisfied $\ell>-1$ \cite{wang2021}. For the gravitational perturbation of the static Einstein-Bumblebee black hole, the perturbation equation is the same as that in Schwarzschild case, hence the QNM frequencies of the gravitational field are completely independent of the Lorentz-violating parameter $\ell$. We calculated the QNMs for massless scalar perturbation, the electromagnetic perturbation with the range $0<\ell<1$, respectively. Fig. \ref{figs}-\ref{figv} show the QNM frequencies for the scalar field and the electromagnetic field in the complex plane as calculated via matrix method.

\subsubsection{A consistency check for massless scalar perturbation}

Before the calculation of the QNMs of the rotation Einstein-Bumblebee black hole, we consider the special case of scalar field at $\ell=0$, i.e., the Kerr black hole. The QNMs of the Kerr black holes are widely discussed \cite{Konoplya2006,Berti2009,Lin20171,Leaver1985}. We compute the QNMs by slow rotation approximation at first and second order, and then compare the results with the exact Kerr results provided by \cite{Berti2009}.

\begin{figure}[h]
\centering
\includegraphics[width=1.0\linewidth]{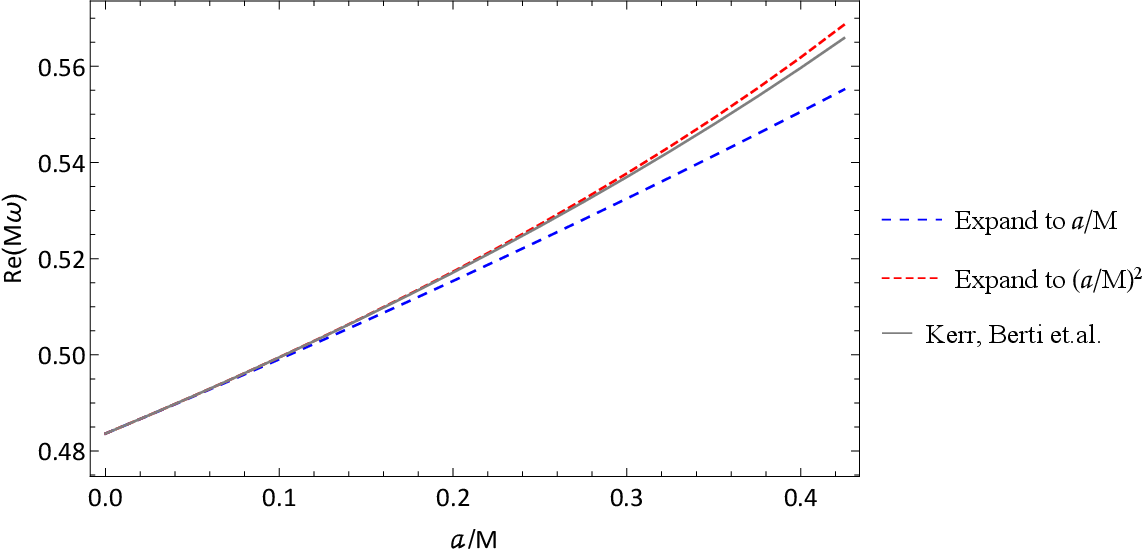}
\includegraphics[width=1.0\linewidth]{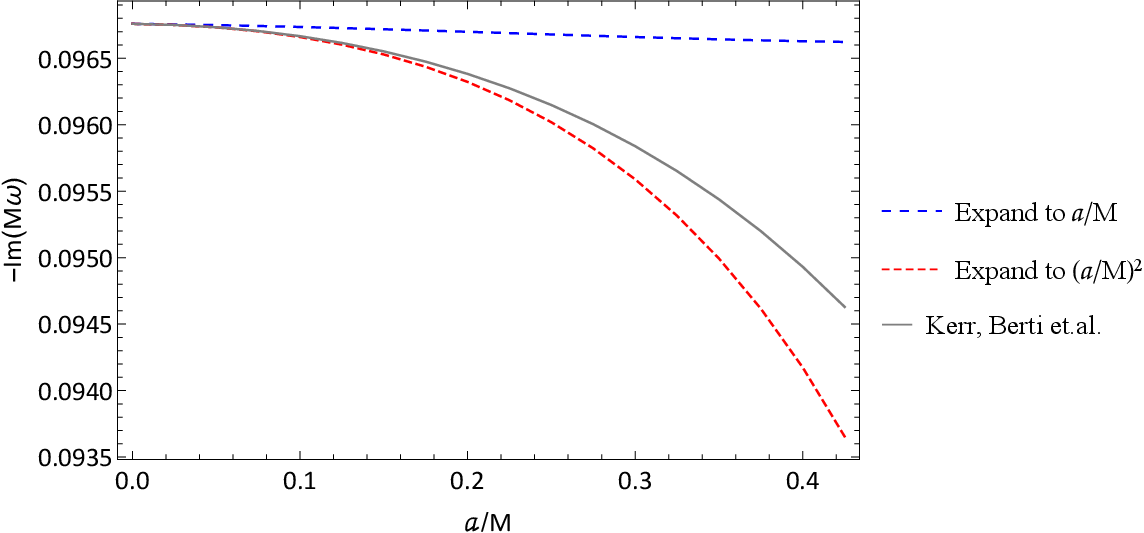}
\caption{Comparison between the exact Kerr result and the results obtained by slow rotation at first or second order for varying values of $a$ with $l=m=2$ and $M=1$.}
\label{figKerr}
\end{figure}

Fig. \ref{figKerr} show that, the QNM frequencies results calculated via slow rotation approximation matches the exact massless scalar perturbation of a Kerr black hole, for $l=m=2$ mode. The figures reveal that, including the real and imaginary parts, the difference between our approach and the exact values is about $1\%$ up to $\tilde{a}=a/M\sim 0.4$. It is obvious that the first order approximation for the QNM frequencies begins to fail at smaller values of $\tilde{a}$, and the second order approximation has much higher accuracy. And it also shows that the frequencies of QNMs do not deviate significantly from the standard result until $\tilde{a}\sim0.2$.

\subsubsection{QNMs of Slowly rotating Einstein-bumblebee black hole}

Now we turn to the slowly rotating Einstein-Bumblebee case with non-vanishing $\tilde{a}$ and $\ell$. The rotation parameter $\tilde{a}$ is limited in the range $0\leq\tilde{a}\leq0.3$.

\begin{figure}[h]
\centering
\includegraphics[width=1.0\linewidth]{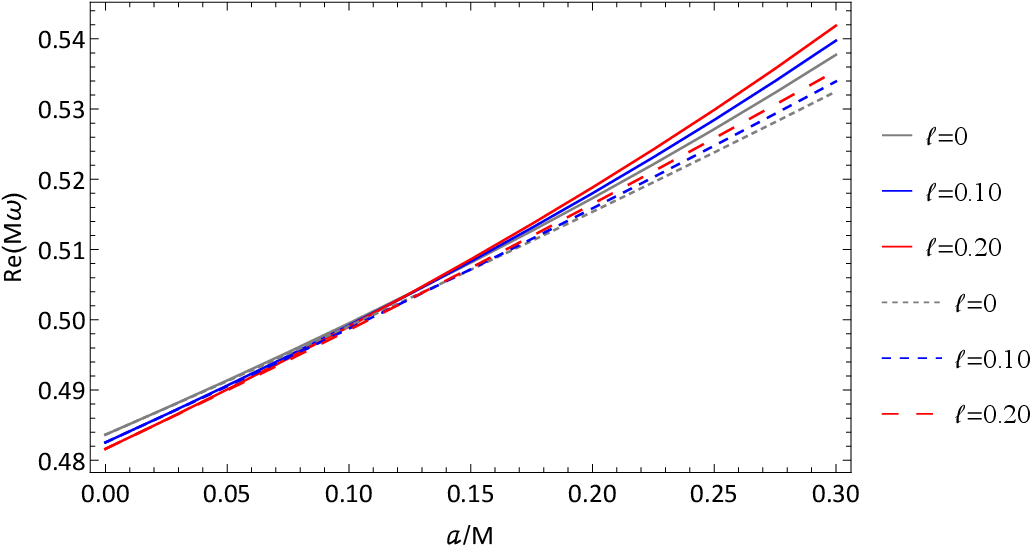}
\includegraphics[width=1.0\linewidth]{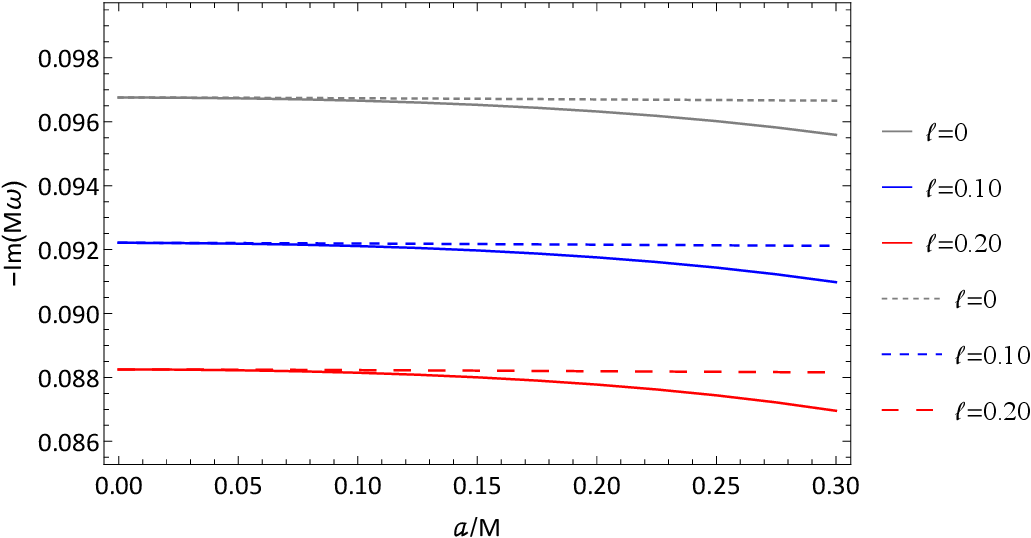}
\caption{Real and imaginary parts of the $ l=m=2 $ massless scalar mode for varying values of $ a $, where the dashed line represents the first-order approximation of the rotation parameter and the solid line represents the second-order approximation of the rotation parameter.}
\label{fig2}
\end{figure}

\begin{figure}[h]
\centering
\includegraphics[width=1.0\linewidth]{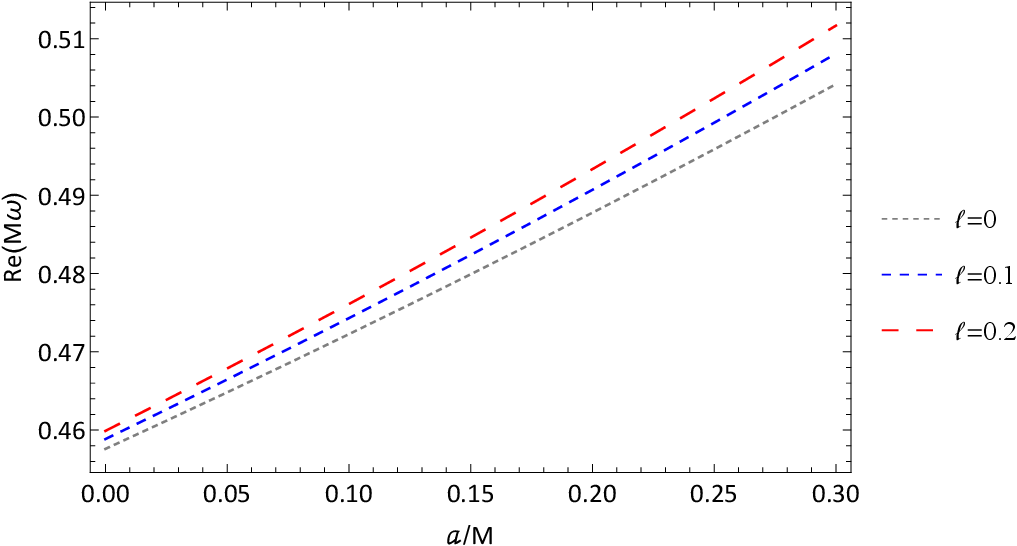}
\includegraphics[width=1.0\linewidth]{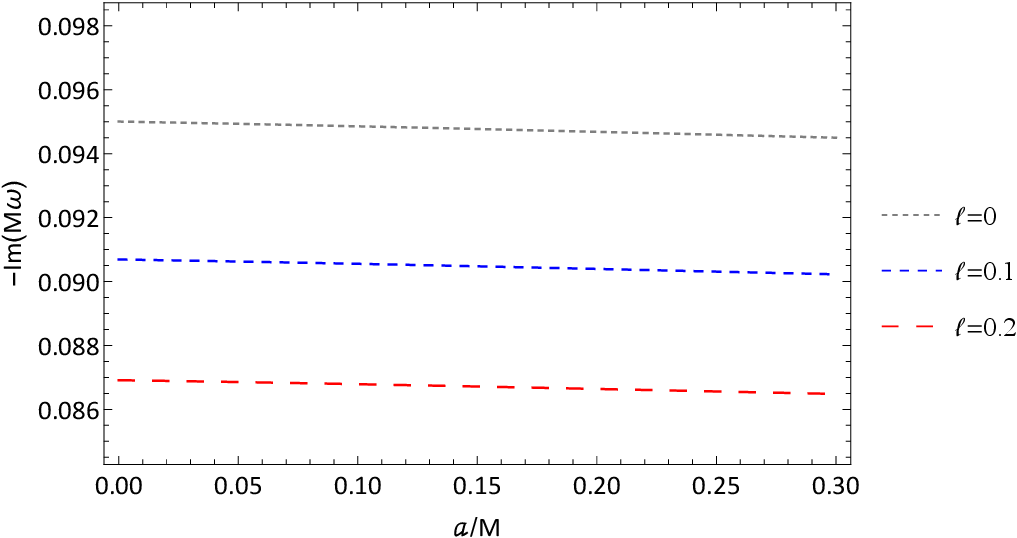}
\caption{Real and imaginary parts of the $l=m=2$ massless electromagnetic mode for varying values of $ a/M $ that belong to the axial sector.}
\label{figv1}
\end{figure}

\begin{figure}[h]
\centering
\includegraphics[width=1.0\linewidth]{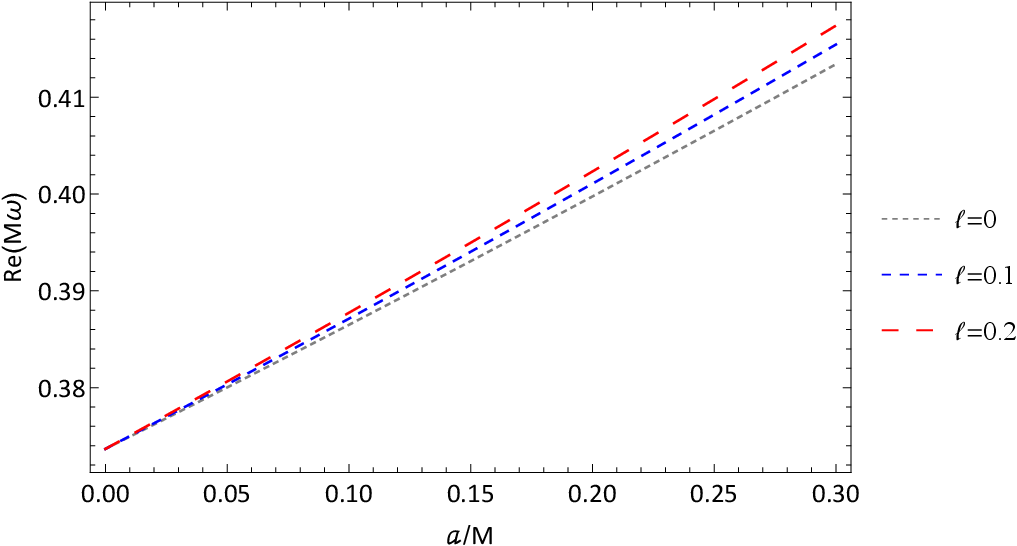}
\includegraphics[width=1.0\linewidth]{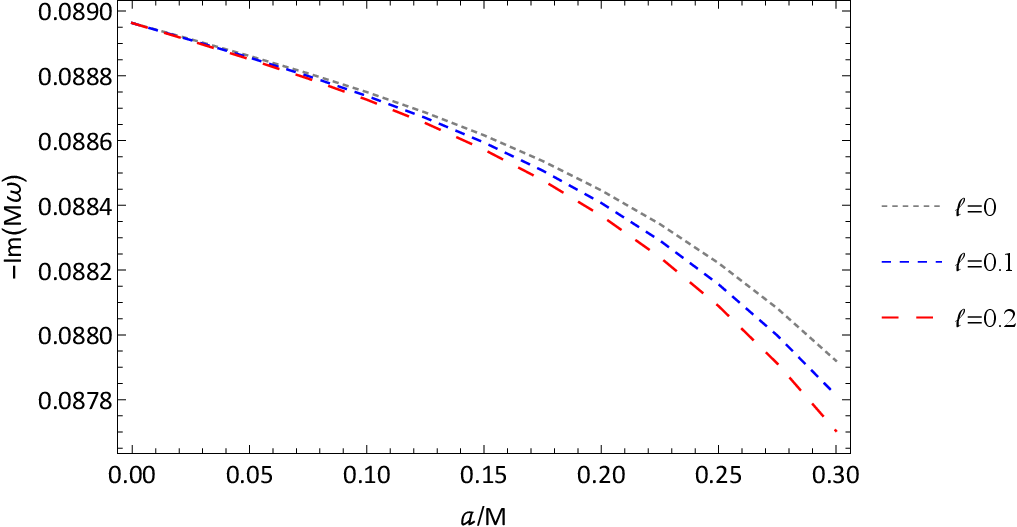}
\caption{Real and imaginary parts of the $l=m=2$ gravitational mode for varying values of $ a/M $.}
\label{figg1}
\end{figure}

For massless scalar field, we computed the QNM frequencies under both the first order and the second order approximations of the rotation. Fig. \ref{fig2} show that, for determined $\ell$, the imaginary part of the second-order approximation shows a more obvious change with the increase of $\tilde{a}$. While for determined $\tilde{a}$, the absolution of the imaginary part will decrease with the increase of $\ell$.

For the electromagnetic and gravitational fields, we calculate the QNM frequencies under the first order approximation of the rotation parameter $\tilde{a}$. Figs. \ref{figv1} and \ref{figg1} show that, in the rotating Einstein-Bumblebee black hole, with the increase of $\ell$ of the Lorentz-violating parameter, the real part of the QNM frequencies will increase, but the absolution of the imaginary part will decrease.

Fig. \ref{figg1} also shows quite intuitively that with a relatively small rotation parameter, $\tilde{a}<0.1$, the Lorentz-violating parameter can hardly have a significant effect on the QNM frequencies. The reason is that for gravitational fields, the Lorentz-violating parameter only affects the first-order rotational correction term of the effective potential. Analyzing the effective potential of the gravitational field shows that following the definition
\begin{align}
\bar{a}=\sqrt{1+\ell}a,
\end{align}
the form of the gravitational field master Eq. \meq{gravV} will remain the same as in the Kerr case \cite{Pani2013IJMPA}. It should be noted that this equivalence is a coincidence at the first order of $\tilde{a}$. Whether this coincidence would be satisfied at the higher order needs further exploration.

The results of QNMs for gravitational perturbations also show that, the parameters $\tilde{a}$ and $\ell$ affecting the QNMs are degenerate. This implies that it is hard to determine whether the deviation of QNMs is coming from the rotation or the LV, and the effect of the bumblebee field may be misinterpreted as being due to the rotation.

\section{CONCLUSIONS AND EXTENSIONS}\label{Sec7}

In this paper, we investigate the calculations of QNMs of the rotation Einstein-Bumblebee black hole, which is an approximation solution of stationary and axisymmetric black hole of Einstein-Bumblebee theory. In this solution, the bumblebee field only has a pure radial vacuum energy expectation and assumed to be as $b_{\mu}=(0,b_r,0,0)$. The strength of the LV can be determined by a coupling constant $\ell$. For the rotation solution, expanding the metric to the second order of rotation parameter $\tilde{a}$, we obtain the first and second order differential equations for the massive scalar perturbation. And we also give the Proca field perturbation equation and the odd parity of gravitational perturbation equation at the first order of $\tilde{a}$, respectively. It shows that the first order scalar field equation and the first order Proca field equation can be written in a form that includes these two perturbations.

Using the matrix method and the continued fraction method, we calculated the QNMs for different fields. Our results show that, the QNMs errors obtained by the two methods are negligible. It clearly show that, as we expected, compare with the exactly result in Kerr solution, the second order approximation has a higher accuracy than the first order approximation. For scalar, vector and gravitational perturbation, we find that the real part of the QNM frequencies are not sensitive to the change of Lorentz-violating parameter $\ell$. However, the results shows that with the increase of $\ell$, the absolute values of the imaginary part of the QNM frequencies decrease, which indicate that the perturbation will decay slower.

It should be noted that the rotation metric we consider corresponding to the bumblebee field has only the radial component. In this theory the Lorentz-violating parameter $\ell$ is determined by both coupling constant $\varrho$ and non-zero VEV $b^a$. Actually, the bumblebee field can be assumed to be $b_{\mu}=(0,b(r),b(\theta),0)$ \cite{chen2020JHEP} or $b_{\mu}=(b(t),b(r),0,0)$ \cite{shao2022}, and the corresponding metrics can be obtained. In these cases, how LV affects the QNMs is also a open problem.

Another point worth noting is that the isospectrality of axial and polar perturbations has been shown in many spacetimes, for example, the Schwarzschild metric and the Reissner Nordstr{\"o}m metric \cite{Chandbook}. The numerical results of Pani et al. provide a strong evidence that the electromagnetic and the gravitational modes of slowly rotation Kerr metric are isospectral at the first order of $\tilde{a}$ \cite{Pani2013prd}. However, the isospectrality is violated in some modified theories of gravity, the violation of isospectrality, such as Lovelock gravity \cite{isoLovelock}, Chern-Simons gravity \cite{isoChSi} and loop quantum gravity \cite{isoLoop}. For Einstein-bumblebee gravity, since we do not know how to construct the master equation for even parity of electromagnetic or gravitational perturbations, whether the bumblebee black hole has the isospectral property is still an open question.

\section{ACKNOWLEDGMENTS}
We would like very much to thank Mengjie Wang and Yunhe Lei for valuable discussions. This work was partially supported by the National Natural Science Foundation of China under Grants No. 11705053, No. 12035005, No. 12122504 and the Hunan Provincial Natural Science Foundation of China under Grant No. 2022JJ40262.

\appendix
\section{Conditions for solutions up to second order of $\tilde{a}$}\label{AppendixA}

In this appendix, we discuss that the metric \meq{metrica2} is the slowly approximation solution of the Einstein-Bumblebee theory, i.e., satisfied the condition \eq{fieldeq}.

By resolving the Einstein-bumblebee gravitational field equations, Ding et. al. presented an Kerr-like solution. But Ref. \cite{Kanzi2022} pointed out that their solution is not the exact solution. However, with sufficiently small Lorentz-violating parameter $\ell$, considering the slow rotation approximation and the parameter $\tilde{a}$ is small, we can get an approximation solution. Substituting the metric \meq{metrica2} into the \eq{fieldeq}, we find the zero components include
\begin{align}
\bar{R}_{tr}=\bar{R}_{t\theta}=\bar{R}_{t\varphi}=\bar{R}_{r\theta}=\bar{R}_{r\varphi}=\bar{R}_{\theta\varphi}=0,
\end{align}
and the non-zero components are
\begin{align}
\bar{R}_{tt}&=\frac{r-2M}{r}\delta,&&\bar{R}_{rr}=\frac{r(1+\ell)}{r-2M}\delta,\non
\bar{R}_{\theta\theta}&=-r^2\delta,&&\bar{R}_{\varphi\varphi}=-r^2\sin^2\theta~\delta,
\end{align}
where $\delta$ is
\begin{align}
\delta=\frac{a^2\ell^2}{4r^4}\left(1+3\cos2\theta\right).
\end{align}
If $\ell$ and $\tilde{a}$ are small parameters, we consider that the non vanishing $\bar{R}_{tt}$ and $\bar{R}_{\theta\theta}$ can be neglected. Hence the metric \meq{metrica2} can be seen as an approximate slowly rotating solution. In our numerical calculation, we restrict the values of $\ell$ and $\tilde{a}$ to $[0,~0.2]$ and $[0,~0.3]$, respectively, to ensure that $\delta\approx0$.

\section{Coefficients of the scalar perturbation}\label{AppendixB}
The coefficients appearing in equation \meq{AlDl} are listed in this Appendix and the relation $\Lambda=l(l+1)$ is also used in this appendix. The coefficients are as follows:
\begin{align}
A_l=
&a^2\left(\frac{4+\Lambda+r^2\left(\mu^2-\omega^2\right)}{2r^4(r-2M)}-\frac{M\left(4+\Lambda+r^2\mu^2\right)}{r^5(r-2M)}
\right.\non
&\left.
+\frac{m^2(1+\ell)}{r^4(r-2M)}-\frac{4M^2(1+\ell)\omega^2}{r^3(r-2M)^2}-\frac{3(r-3M)}{r^6(1+\ell)}
\right)\Psi_l\non
&+\frac{2M\left(\Lambda+r^2\mu^2\right)}{r^3(r-2M)}\Psi_l-\frac{\Lambda+r^2\left(\mu^2-\omega^2\right)}{r^2(r-2M)}\Psi_l\non
&+\frac{2M}{r^4(1+\ell)}\left(r\Psi'_l-\Psi_l\right)+\frac{r-2M}{r^2(1+\ell)}\Psi''_l\non
&+\frac{a^2(r+2M+2r\ell)}{2r^4(1+\ell)}\Psi''_l-\frac{a^2(5M+2r\ell)}{r^5(1+\ell)}\Psi'_l\non
&-\frac{4mMa\sqrt{1+\ell}\omega}{r^3(r-2M)}\Psi_l,
\end{align}
\begin{align}
D_l=&\frac{2M}{r^6}\Psi_l-\frac{(1+\ell)\left[2M\left(\Lambda+r^2\omega^2\right)-r\Lambda\right]}
{r^5(r-2M)}\Psi_l\non
&-\frac{2M}{r^5}\Psi_l'-\frac{r-2M}{r^4}\Psi_l''.
\end{align}

\section{Coefficients and source terms of the Proca perturbation}\label{AppendixC}
In this appendix we present the explicit expressions of the parameters and the relations for the massive vector perturbation. The explicit form of the coefficients appearing in Eqs. \meq{group1}-\meq{group3} is
\begin{align}
A^{(0)}_l=&\frac{i(r-4M)\omega}{r^2(r-2M)(1+\ell)}\mathrm{R}_l-\frac{\Lambda+r^2\mu^2}{r^3}\mathrm{P}_l-\frac{i\omega}{r^2}\mathrm{S}_l\non
&+\frac{i\omega}{r(1+\ell)}\mathrm{R}'_l+\frac{r-2M}{r^2(1+\ell)}\mathrm{P}''_l+\frac{2mMa\sqrt{1+\ell}}{r^3(r-2M)}\non
&\times\left[\frac{i}{r(1+\ell)}\left(\mathrm{R}_l-\frac{r-2M}{\Lambda}\mathrm{S}'_l \right)
-\omega\mathrm{P}_l-\frac{ir\omega^2}{\Lambda}\mathrm{S}_l \right]\non
&+\frac{m^2a^2(1+\ell)}{r^4(r-2M)}\left(\mathrm{P}_l+\frac{i\omega r}{\Lambda}\mathrm{S}_l\right)
+\frac{ia^2\omega}{r^3(1+\ell)}\mathrm{R}'_l\non
&+\frac{2a^2(r+M)}{r^6}\left(\mathrm{P}_l-r\mathrm{P}'_l+\frac{r^3}{2(r+M)}\mathrm{P}''_l\right)\non
&+\frac{i\omega a^2}{r-2M}\left(\frac{4M^2}{r^4(r-2M)} -\frac{1}{r^3(1+\ell)}\right)\mathrm{R}_l,
\end{align}
\begin{align}
A^{(1)}_l=&\frac{2M\left(\Lambda+r^2\mu^2\right)-r\Lambda-r^3\left(\mu^2-\omega^2\right)}{r^2(r-2M)^2}\mathrm{R}_l+\frac{1}{r^2}\mathrm{S}'_l\non
&+\frac{i \omega}{r(r-2M)}\mathrm{P}_l-\frac{i\omega}{r-2M}\mathrm{P}'_l-\frac{2iaMm\sqrt{1+\ell}}{r^2(r-2M)\Lambda}\non
&\times\left[\frac{\Lambda}{r^2}\mathrm{P}_l-\frac{2i\Lambda\omega}{r-2M}\mathrm{R}_l-\frac{\Lambda}{r}\mathrm{P}'_l+i\omega\mathrm{S}'_l  \right]\non
&+a^2\left(\frac{r\omega^2-\mu^2(r-2M)}{r^2(r-2M)^2}-\frac{\Lambda}{r^4(r-2M)} \right)\mathrm{R}_l\non
&+\frac{(1+\ell)a^2}{r^4(r-2M)^2}\left[\left(r\Lambda-\frac{4r^2M^2+r^4}{r-2M} \right)\mathrm{R}_l\right.\non
&+\left.\left(rm^2+r^3\mu^2\right)\mathrm{R}_l+4iM^2\omega\left(r\mathrm{P}'_l-\mathrm{P}_l\right)
\right.\non
&\left.-\frac{r(r-2M)m^2}{\Lambda}\mathrm{S}'_l
\right]
\end{align}
\begin{align}
A^{(L)}_l=
&-\frac{2aMm\sqrt{1+\ell}}{\Lambda(r-2M)r^3}\left(i\Lambda\mathrm{P}_l+r\omega\mathrm{S}_l \right)
+\frac{a^2(1+\ell)}{r^3(r-2M)}\non
&\times\left(\frac{m^2}{\Lambda}\mathrm{S}_l-\frac{4iM^2\omega}{r-2M}\mathrm{P}_l\right)+\frac{a^2}{r^4(1+\ell)}\left(r\mathrm{R}'_l-\mathrm{R}_l\right)\non
&+\frac{i \omega}{r-2M}\mathrm{P}_l+\frac{\mathrm{R}_l+r\mathrm{R}'_l}{r^2(1+\ell)}-\frac{1}{r^2}\mathrm{S}_l,
\end{align}
\begin{align}
\tilde{A}^{(0)}_l=&\frac{4aM\sqrt{1+\ell}}{r^5}\mathrm{Q}_l-\frac{4a^2Mm(1+\ell )\omega}{\Lambda r^5}\mathrm{Q}_l,\\
\tilde{A}^{(1)}_l=&-\frac{2ia^2m(1+\ell)}{\Lambda r^4}\mathrm{Q}'_l,\\
\tilde{A}^{(L)}_l=&0,
\end{align}
\begin{align}
B^{(0)}_l=&\frac{2aM\sqrt{1+\ell}\omega^2}{r^2(r-2M)\Lambda}\mathrm{Q}_l+\frac{2aM}{r^4\sqrt{1+\ell}\Lambda}\mathrm{Q}'_l\non
&-\frac{a^2m(1+\ell)\omega}{r^3(r-2M)\Lambda}\mathrm{Q}_l,
\end{align}
\begin{align}
B^{(1)}_l=&\frac{2i aM\sqrt{1+\ell}\omega}{\Lambda(r-2M)r^2}\mathrm{Q}'_l-\frac{ia^2m(1+\ell)}{\Lambda(r-2M)r^3}\mathrm{Q}'_l,
\end{align}
\begin{align}
B^{(L)}=&-\frac{2iaM\sqrt{1+\ell}\omega}{\Lambda(r-2M)r^2}\mathrm{Q}_l+\frac{ia^2m(1+\ell)}{\Lambda(r-2M)r^3}\mathrm{Q},
\end{align}
\begin{align}
\tilde{B}^{(0)}_l=&\frac{4a^2M(1+\ell)\left(\Lambda\mathrm{P}_l+ir\omega\mathrm{S}_l\right)}{\Lambda r^6},\\
\tilde{B}^{(1)}_l=&\frac{2a^2(1+\ell)\left[\Lambda\mathrm{R}_l-(r-2M)\mathrm{S}'_l\right]}{\Lambda(r-2M)r^4},\\
\tilde{B}^{(L)}_l=&0,
\end{align}
\begin{align}
D^{(0)}_l=&\frac{a^2}{(r-2M)r^6}\left[(2M-r)\left(6M-\Lambda r(1+\ell) \right)\mathrm{P}_l\right.\non
&\left.+r\left(
i(10M-r)r\omega \mathrm{R}_l-(2M-r)\left(ir (1+\ell)\omega\mathrm{S}_l\right.\right.\right.\non
&\left.\left.\left.
+6M\mathrm{P}'_l+r\left( -ir\omega \mathrm{R}'_l+(2M-r)\mathrm{P}''_l \right)\right)\right) \right],
\end{align}
\begin{align}
D^{(1)}_l=&\frac{a^2(1+\ell)}{r^4(r-2M)^2}\left[\left( \Lambda(r-2M)-2Mr^2\omega^2 \right)\mathrm{R}_l\right.\non
&(r-2M)\left(2iMr\omega\mathrm{P}'_l-(r-2M)\mathrm{S}'_l \right)\non
&\left.+2iM(2M-r)\omega\mathrm{P}_l  \right],\\
D^{(L)}_l=&-\frac{a^2}{(r-2M)r^4}\left[2iMr(1+\ell)\omega\mathrm{P}_l\right.\non
&\left.+(r-2M)\left(\mathrm{R}_l-(1+\ell)\mathrm{S}_l+r\mathrm{R}'_l \right)  \right],
\end{align}
\begin{align}
\alpha_l=&\frac{(1+\ell)^{-1}}{(2M-r)r^2\Lambda}\big[ir^2(1+\ell)\Lambda \omega\mathrm{P}_l-(r-2M)\Lambda\mathrm{R}_l\non
& +r^2(1+\ell)\left(r\mu^2-r\omega^2-2M\mu^2 \right)\mathrm{S}_l+r\Lambda(r-2M)\mathrm{R}'_l \non
& -(r-2M)\left(2M\mathrm{S}'_l+r(r-2M)\mathrm{S}''_l \right) \big] \non
&+\frac{2aMm\sqrt{1+\ell}}{\Lambda(r-2M)r^3}\left(i\Lambda \mathrm{P}_l-r\omega\mathrm{S}_l \right)
+\frac{a^2(1+\ell)^{-1}}{r^4(r-2M)^2\Lambda}\non
&\times\left[ 2i\Lambda Mr^2(1+\ell)^2\omega\mathrm{P}_l-3\Lambda(r-2M)^2\ell\mathrm{R}_l\right.\non
&\left.+(r-2M)^2\left(\Lambda r\ell \mathrm{R}'_l-2M(1+\ell)\left(3\mathrm{S}'_l-r\mathrm{S}''_l \right) \right)\right.\non
&\left.-2Mr^3(1+\ell)^2\omega^2\mathrm{S}_l \right],
\end{align}
\begin{align}
\beta_l=&\left(\frac{1}{r^2}+\frac{\mu^2}{\Lambda}-\frac{r\omega^2}{\Lambda(r-2M)} \right)\mathrm{Q}_l-\frac{2M}{r^2\Lambda (1+\ell)}\mathrm{Q}'_l\non
&-\frac{(r-2M)}{r\Lambda(1+\ell)}\mathrm{Q}''_l+\frac{2aMm\sqrt{1+\ell}\omega}{\Lambda(r-2M)r^2}\mathrm{Q}_l\non
&-\frac{(1+\ell)a^2}{r^4}\mathrm{Q}_l+\frac{2Ma^2(1+\ell)\omega^2}{r(r-2M)^2\Lambda}\mathrm{Q}_l\non
&+\frac{2Ma^2}{r^3\Lambda}\left(\frac{3}{r}\mathrm{Q}'_l-\mathrm{Q}''_l\right),
\end{align}
\begin{align}
\zeta_l=&\frac{6aM}{r^4\sqrt{1+\ell}}\left(\frac{ir^2\omega}{r-2M}\mathrm{R}_l-\mathrm{P}_l+r\mathrm{P}'_l \right)+\frac{a^2m}{r^4\Lambda}\non
&\times\left[\frac{2Mr(1+\ell)\Lambda\omega}{r-2M} \mathrm{P}_l-\frac{i(8M-r)\Lambda}{r-2M}\mathrm{R}_l-ir\Lambda\mathrm{R}'_l\right.\non
&\left.+\frac{2iMr^2(1+\ell)\omega^2}{r-2M}\mathrm{S}_l+i\left(8M\mathrm{S}'_l+(r-2M)r\mathrm{S}''_l\right)\right],
\end{align}
\begin{align}
\eta_l=&\frac{2iaM\sqrt{1+\ell}\omega}{(2M-r)r^2}\mathrm{Q}_l+\frac{ia^2m}{r^4\Lambda}\bigg[
\frac{2M(1+\ell)\left(\Lambda+r^2\omega^2\right)}{r-2M}\mathrm{Q}_l\non
&-2\left(r-3M\right)\mathrm{Q}'_l+r(r-2M)\mathrm{Q}''_l\bigg],
\end{align}
\begin{align}
\rho_l=&\frac{a^2}{r^4(r-2M)\Lambda}\left[2Mr(1+\ell)\omega\left(r\omega\mathrm{S}_l-i\Lambda\mathrm{P}_l\right)\right.\non
&\left.-2(r-3M)(r-2M)\mathrm{S}'_l+r(r-2M)^2\mathrm{S}''_l\right.\non
&\left.+\Lambda(r-2M)\left(3\mathrm{R}_l-r\mathrm{R}'_l\right)
\right],
\end{align}
\begin{align}
\lambda_l=&-\frac{4a^2M(1+\ell)}{r^5}\mathrm{Q}_l,\\
\sigma_l=&0,
\end{align}
\begin{align}
\gamma_l=&\frac{a^2}{r^4(r-2M)\Lambda}\left[(1+\ell)\left(r\Lambda-2M(\Lambda+r^2\omega^2)\right)\mathrm{Q}_l\right.\non
&\left.-8M(r-2M)\mathrm{Q}'_l-r(r-2M)^2\mathrm{Q}''_l  \right].
\end{align}

After the angular parts are separated, \eqs{group1} to \meq{group3} can be rewritten as the pure radial equations, which are given by
\begin{align}\label{MEQtrL}
&A^{(I)}_l+\mathcal{Q}^2_{l+1}\left[D^{(I)}_l+l\tilde{B}^{(I)}_l \right]
+\mathcal{Q}^2_l\left[D^{(I)}_l-(l+1)\tilde{B}^{(I)}_l\right]\non
&+\mathcal{Q}_l\left[\tilde{A}^{(I)}_{l-1}+(l-1)B^{I}_{l-1} \right]+\mathcal{Q}_{l+1}\left[\tilde{A}^{(I)}_{l+1}-(l+2)B^{(I)}_{l+1} \right]\non
&+\mathcal{Q}_{l-1}\mathcal{Q}_l\left[D^{(I)}_{l-2}+(l-2)\tilde{B}^{(I)}_{l-2} \right]\non
&+\mathcal{Q}_{l+2}\mathcal{Q}_{l+1}\left[D^{(I)}_{l+2}-(l+3)\tilde{B}^{(I)}_{l+2} \right]=0,
\end{align}
\begin{align}\label{MEQtheta}
&\mathcal{Q}_{l+1}\left[l \eta_{l+1}+i\hsp m\left((l+2)\gamma_{l+1}-\lambda_{l+1}\right) \right]\non
&-\mathcal{Q}_{l-1}\mathcal{Q}_l(l+1)\left[(l-2)\rho_{l-2}+\sigma_{l-2} \right]\non
&+\mathcal{Q}_{l+2}\mathcal{Q}_{l+1}l\left[-(l+3)\rho_{l+2}+\sigma_{l+2} \right]\non
&+\Lambda\alpha_l-i\hsp m\zeta_l+\mathcal{Q}^2_{l+1}l\left[l\rho_l+\sigma_l \right]\non
&+\mathcal{Q}^2_l(l+1)\left[(l+1)\rho_l-\sigma_l\right]\non
&+\mathcal{Q}_l\left[-(l+1)\eta_{l-1}-i\hsp  m\left((l-1)\gamma_{l-1}+\lambda_{l-1}\right) \right]=0,
\end{align}
\begin{align}\label{MEQvarphi}
&\mathcal{Q}_{l+1}\left[l \zeta_{l+1}-i\hsp m\left((l+2)\rho_{l+1}-\sigma_{l+1}\right) \right]\non
&-\mathcal{Q}_{l-1}\mathcal{Q}_l(l+1)\left[(l-2)\gamma_{l-2}+\lambda_{l-2} \right]\non
&+\mathcal{Q}_{l+2}\mathcal{Q}_{l+1}l\left[-(l+3)\gamma_{l+2}+\lambda_{l+2} \right]\non
&+\Lambda\beta_l+i\hsp m\eta_l+\mathcal{Q}^2_{l+1}l\left[l\gamma_l+\lambda_l \right]\non
&+\mathcal{Q}^2_l(l+1)\left[(l+1)\gamma_l-\lambda_l\right]\non
&+\mathcal{Q}_l\left[-(l+1)\zeta_{l-1}+i\hsp  m\left((l-1)\rho_{l-1}+\sigma_{l-1}\right) \right]=0.
\end{align}

The source term corresponding to the coupled system described by Eqs. \meq{D2Ra1}-\meq{D2Sa1} is
\begin{align}
\mathcal{S}^{\text{po}}_1=&(1+\ell)^{\frac{3}{2}}\frac{6 i a\hsp M\mathcal{F}\omega}{\Lambda r^3}
\left[l\mathcal{Q}_{l+1}\mathrm{Q}_{l+1}-(l+1)\mathcal{Q}_l\mathrm{Q}_{l-1} \right]\non
&+\sqrt{1+\ell}
\frac{2 a\hsp M m}{\Lambda r^5\omega}\left[\Lambda\left(2(1+\ell)r^2\omega^2+3 \mathcal{F}^2\right)
\mathrm{R}_l\right.\non
&\left.+3\mathcal{F}\left(r \Lambda \mathcal{F}~(\mathrm{R}_l)'-(1+\ell)\left(\Lambda \mathcal{F}+r^2\omega^2\right)\mathrm{S}_l \right)
\right],
\\
\mathcal{S}^{\text{po}}_2=&\sqrt{1+\ell}\frac{2a\hsp
Mm}{r^5\omega}\left[2(1+\ell)r^2\omega^2\mathrm{S}_l+3r\mathcal{F}^2\mathrm{S}_l'\right.\non
&\left.-3\mathcal{F}\left(\Lambda+r^2\mu^2 \right)\mathrm{R}_l \right],
\end{align}
and the corresponding source term of Eq. \meq{D2Qa1} is
\begin{align}
\mathcal{S}^{\text{ax}}=
-\frac{6i\hsp aM\mathcal{F}}{\sqrt{1+\ell}r^5\omega}\left[(l+1)\mathcal{Q}_l\psi_{l-1}-l\mathcal{Q}_{l+1}\psi_{l+1} \right],
\end{align}
where we have defined the polar function
\begin{align}
\psi_l=(1+\ell)\left[(\Lambda+r^2\mu^2)\mathrm{R}_l-(r-2M)\mathrm{S}_l'\right].
\end{align}
Note that this term is similar to Eq. (25) in Ref. \cite{Rosa2012}. As expected, the axial perturbation $ \mathrm{Q}_l $ is coupled to the polar functions with $ l\pm1 $.

\section{recurrence coefficients of gravitational Perturbation}\label{AppendixD}

In this appendix, we present the explicit expressions of the recurrence coefficients when we used the continued fraction method. The recurrence coefficients in Eq. \meq{grav0-4} are
\begin{align}
n_1\mathcal{C}_{1,0}=&\tilde{\ell}m\tilde{a}[3i+\Lambda\omega(1-3i\omega)]\non
&-i\Lambda\omega[\Lambda-3-4\omega(i+2\omega)],\\
n_2\mathcal{C}_{2,0}=&\tilde{\ell}m\tilde{a}[30i+\Lambda(1-2i\omega)\omega]\non
&i\Lambda\omega[3+4\omega(i+\omega)],\\
n_2\mathcal{C}_{2,1}=&3i\tilde{\ell}m\tilde{a}[\Lambda\omega(i+\omega)-1]\non
&+i\Lambda\omega[1+\Lambda-4\omega(3i+2\omega)],
\end{align}
\begin{align}
n_3\mathcal{C}_{3,0}=&72i\tilde{\ell}m\tilde{a},\\
n_3\mathcal{C}_{3,1}=&4\Lambda(2-i\omega)\omega^2+2i\tilde{\ell}m\tilde{a}[\Lambda\omega(i+\omega)-15],\\
n_3\mathcal{C}_{3,2}=&\tilde{\ell}m\tilde{a}[3i+\Lambda(5-3i\omega)\omega]\non
&-i\Lambda\omega[9+\Lambda-4\omega(5i+2\omega)],
\end{align}
\begin{align}
n_4\mathcal{C}_{4,0}=&66i\tilde{\ell}m\tilde{a},\\
n_4\mathcal{C}_{4,1}=&-72i\tilde{\ell}m\tilde{a},\\
n_4\mathcal{C}_{4,2}=&\tilde{\ell}m\tilde{a}[30i+\Lambda(3-2i\omega)\omega]\non
&+i\Lambda\omega[4\omega(3i+\omega)-5],\\
n_4\mathcal{C}_{4,3}=&i\Lambda\omega[21+\Lambda-4\omega(7i+2\omega)]\non
&+i\tilde{\ell}m\tilde{a}[\Lambda\omega(7i+3\omega)-3],
\end{align}
where
\begin{align}
n_1=&\Lambda\omega(\tilde{\ell}m\tilde{a}-2\omega-i),\\
n_2=&2\Lambda\omega(2i-\tilde{\ell}m\tilde{a}+2\omega),\\
n_3=&3\Lambda\omega(\tilde{\ell}m\tilde{a}-2\omega-3i),\\
n_4=&4\Lambda\omega(4i-\tilde{\ell}m\tilde{a}+2\omega).
\end{align}
The recurrence coefficients in Eq. \meq{6oder} are
\begin{align}
\alpha_n=&4\Lambda(1+n)(1+n+i\tilde{\ell}m\tilde{a}-2i\omega)\omega,\\
\beta_m=&4\Lambda\omega[3-\Lambda-2n^2-n(2-8i\omega)+4i\omega+8\omega^2]\non
&+4\tilde{\ell}m\tilde{a}[3-\Lambda\omega(i+2in+3\omega)],\\
\gamma_m=&4\Lambda\omega(n^2-4-4in\omega-4\omega^2)\non
&+4\tilde{\ell}m\tilde{a}[i\Lambda n \omega+2(\Lambda\omega^2-15)],\\
\sigma_n=&288\tilde{\ell}m\tilde{a},\\
\tau_n=&-264\tilde{\ell}m\tilde{a},\\
\delta_n=&84\tilde{\ell}m\tilde{a}.
\end{align}
\begin{widetext}

\section{QNM FREQUENCY TABLES}\label{AppendixE}

\begin{table}[H]

\renewcommand{\arraystretch}{1.2}
\centering
\caption{Comparison of the $n=0$, $l=m=2$ mode massless scalar QNM frequencies calculated by the matrix method and the continued fraction method under the first order slow rotation approximation.}

\label{tab:1}
\setlength\tabcolsep{5mm}{
\begin{tabular}{cccccccc}
\hline\hline
\multirow{2}{*}{~\bf }	&
\multirow{3}{*}{~\bf $ a $~}  &
\multicolumn{2}{c}{Matrix method} & \multicolumn{2}{c}{Continued fraction method} & \multicolumn{2}{c}{\% error } \\
\cline{3-4}  \cline{5-6} \cline{7-8}
&~&{Re$(M\omega)$ }&{-Im$(M\omega)$}&{Re$(M\omega)$}&{-Im$(M\omega)$}&{Re$(M\omega)$}&{-Im$(M\omega)$}\\
\hline
\multirow{5}{*}{$ \ell=0 $}
&0     &0.483644 &0.096759 &0.483644 &0.096759 &0.000044 &0.000027 \\
&0.05  &0.491268 &0.096749 &0.491268 &0.096749 &0.000042 &0.000059 \\
&0.10  &0.499093 &0.096735 &0.499093 &0.096735 &0.000040 &0.000089 \\
&0.15  &0.507125 &0.096718 &0.507125 &0.096718 &0.000037 &0.000118 \\
&0.20  &0.515367 &0.096699 &0.515367 &0.096699 &0.000033 &0.000143 \\
\\
\multirow{5}{*}{$ \ell=0.1 $}
  &0     &0.482542 	&0.092212 	&0.482542 	&0.092212 	&0.000032 	&0.000026
\\&0.05  &0.490533 	&0.092203 	&0.490533 	&0.092203 	&0.000031 	&0.000053
\\&0.10  &0.498746 	&0.092189 	&0.498746 	&0.092189 	&0.000029 	&0.000078
\\&0.15  &0.507188 	&0.092172 	&0.507188 	&0.092172 	&0.000026 	&0.000101
\\&0.20  &0.515863 	&0.092153 	&0.515863 	&0.092153 	&0.000023 	&0.000122
\\
\\
\multirow{5}{*}{$ \ell=0.2 $}
  &0     &0.481622 	&0.088251 	&0.481622 	&0.088251 	&0.000024 	&0.000025
\\&0.05  &0.489964 	&0.088242 	&0.489964 	&0.088242 	&0.000023 	&0.000048
\\&0.10  &0.498551 	&0.088229 	&0.498550 	&0.088229 	&0.000021 	&0.000069
\\&0.15  &0.507387 	&0.088212 	&0.507387 	&0.088212 	&0.000019 	&0.000088
\\&0.20  &0.516479 	&0.088193 	&0.516479 	&0.088193 	&0.000016 	&0.000105
\\
\hline\hline
\end{tabular}}
\end{table}

\begin{table}[h]
\renewcommand{\arraystretch}{1.2}
\centering
\caption{Comparison of the $n=0$, $l=m=2$ mode massless electromagnetic QNM frequencies calculated by the matrix method and the continued fraction method under the first order slow rotation approximation.}
\label{tab:2}
\setlength\tabcolsep{5mm}{
\begin{tabular}{cccccccc}
\hline\hline
\multirow{2}{*}{~\bf }	&
\multirow{3}{*}{~\bf $ a $~}  &
\multicolumn{2}{c}{Matrix method} & \multicolumn{2}{c}{Continued fraction method} & \multicolumn{2}{c}{\% error } \\
\cline{3-4}  \cline{5-6} \cline{7-8}
&~&{Re$(M\omega)$ }&{-Im$(M\omega)$}&{Re$(M\omega)$}&{-Im$(M\omega)$}&{Re$(M\omega)$}&{-Im$(M\omega)$}\\
\hline
\multirow{5}{*}{$ \ell=0 $}
  &0     &0.457596 	&0.095005 	&0.457596 	&0.095004 	&0.000052 	&0.000190
\\&0.05  &0.464821 	&0.094935 	&0.464821 	&0.094934 	&0.000045 	&0.000230
\\&0.10  &0.472255 	&0.094857 	&0.472255 	&0.094857 	&0.000037 	&0.000264
\\&0.15  &0.479904 	&0.094773 	&0.479904 	&0.094773 	&0.000028 	&0.000292
\\&0.20  &0.487773 	&0.094684 	&0.487773 	&0.094684 	&0.000019 	&0.000313
\\
\\
\multirow{5}{*}{$ \ell=0.1 $}
  &0     &0.458850 	&0.090690 	&0.458850 	&0.090690 	&0.000038 	&0.000143
\\&0.05  &0.466461 	&0.090626 	&0.466461 	&0.090626 	&0.000032 	&0.000176
\\&0.10  &0.474301 	&0.090554 	&0.474301 	&0.090554 	&0.000026 	&0.000204
\\&0.15  &0.482377 	&0.090476 	&0.482377 	&0.090476 	&0.000019 	&0.000226
\\&0.20  &0.490697 	&0.090394 	&0.490697 	&0.090394 	&0.000012 	&0.000241
\\
\\
\multirow{5}{*}{$ \ell=0.2 $}
  &0     &0.459896 	&0.086913 	&0.459896 	&0.086913 	&0.000028 	&0.000112
\\&0.05  &0.467874 	&0.086855 	&0.467873 	&0.086855 	&0.000023 	&0.000139
\\&0.10  &0.476102 	&0.086788 	&0.476102 	&0.086788 	&0.000018 	&0.000161
\\&0.15  &0.484589 	&0.086715 	&0.484589 	&0.086715 	&0.000013 	&0.000179
\\&0.20  &0.493342 	&0.086639 	&0.493342 	&0.086639 	&0.000007 	&0.000191
\\
\hline\hline
\end{tabular}}
\end{table}

\begin{table}[h]
\renewcommand{\arraystretch}{1.2}
\centering
\caption{Comparison of the $n=0$, $l=m=2$ mode gravitational QNM frequencies calculated by the matrix method and the continued fraction method under the first order slow rotation approximation.}
\label{tab:3}
\setlength\tabcolsep{5mm}{
\begin{tabular}{cccccccc}
\hline\hline
\multirow{2}{*}{~\bf }	&
\multirow{3}{*}{~\bf $ a $~}  &
\multicolumn{2}{c}{Matrix method} & \multicolumn{2}{c}{Continued fraction method} & \multicolumn{2}{c}{\% error } \\
\cline{3-4}  \cline{5-6} \cline{7-8}
&~&{Re$(M\omega)$ }&{-Im$(M\omega)$}&{Re$(M\omega)$}&{-Im$(M\omega)$}&{Re$(M\omega)$}&{-Im$(M\omega)$}\\
\hline
\multirow{5}{*}{$ \ell=0 $}
  &0     &0.373671 	&0.088963 	&0.373670 	&0.088959 	&0.000176 	&0.005109
\\&0.05  &0.380020 	&0.088861 	&0.380019 	&0.088858 	&0.000198 	&0.004226
\\&0.10  &0.386486 	&0.088750 	&0.386486 	&0.088747 	&0.000200 	&0.003370
\\&0.15  &0.393064 	&0.088616 	&0.393063 	&0.088614 	&0.000187 	&0.002555
\\&0.20  &0.399747 	&0.088446 	&0.399746 	&0.088445 	&0.000166 	&0.001789
\\
\\
\multirow{5}{*}{$ \ell=0.1 $}
  &0     &0.373671 	&0.088963 	&0.373670 	&0.088959 	&0.000176 	&0.005109
\\&0.05  &0.380333 	&0.088856 	&0.380332 	&0.088853 	&0.000199 	&0.004183
\\&0.10  &0.387124 	&0.088738 	&0.387123 	&0.088735 	&0.000199 	&0.003289
\\&0.15  &0.394036 	&0.088594 	&0.394035 	&0.088592 	&0.000185 	&0.002439
\\&0.20  &0.401063 	&0.088407 	&0.401063 	&0.088406 	&0.000161 	&0.001646
\\
\\
\multirow{5}{*}{$ \ell=0.2 $}
  &0     &0.373671 	&0.088963 	&0.373670 	&0.088959 	&0.000176 	&0.005109
\\&0.05  &0.380632 	&0.088851 	&0.380631 	&0.088848 	&0.000199 	&0.004142
\\&0.10  &0.387734 	&0.088726 	&0.387733 	&0.088724 	&0.000199 	&0.003211
\\&0.15  &0.394967 	&0.088572 	&0.394966 	&0.088570 	&0.000182 	&0.002330
\\&0.20  &0.402325 	&0.088368 	&0.402324 	&0.088367 	&0.000156 	&0.001511
\\
\hline\hline
\end{tabular}}
\end{table}
\end{widetext}

~\\


\begin{thebibliography}{000}

\bibitem{Kostelecky1998}
D. Colladay D and V. A. Kosteleck, \textit{Lorentz-violating extension of the standard model.} Phys. Rev. D, \textbf{58}, 116002 (1998).

\bibitem{Kostelecky1991}
V. A. Kosteleck and R. Potting, \textit{CPT and strings.} Nucl. Phys. B \textbf{359}, 545 (1991).

\bibitem{Kostelecky19891}
V. A. Kosteleck and S. Samuel, \textit{Phenomenological gravitational constraints on strings and higher-dimensional theories.} Phys. Rev. Lett. \textbf{63}, 224 (1989).

\bibitem{Gambini1999}
R. Gambini and J. Pullin, \textit{Nonstandard optics from quantum space-time.} Phys. Rev. D, \textbf{59}, 124021 (1999).

\bibitem{Kostelecky2001}
S. M. Carroll, J. A. Harvey and V. A. Kosteleck, et al. \textit{Noncommutative field theory and Lorentz violation.} Phys. Rev. Lett. \textbf{87}, 141601 (2001).

\bibitem{Ferrari2007}
A. F. Ferrari and M. Gomes, et al. \textit{Lorentz violation in the linearized gravity}. Phys. Lett. B, \textbf{652}, 174-180 (2007).

\bibitem{Garay1998}
L. J. Garay, \textit{Spacetime foam as a quantum thermal bath.} Phys. Rev. Lett. \textbf{80}, 2508 (1998).

\bibitem{Kostelecky20030}
V. A. Kostelecky, R. Lehnert and M. J. Perry, \textit{Spacetime-varying couplings and Lorentz violation.} Phys. Rev. D \textbf{68}, 123511 (2003).

\bibitem{Bertolami2004}
O. Bertolami, et al. \textit{Cosmological acceleration, varying couplings, and Lorentz breaking.} Phys. Rev. D \textit{69}, 083513 (2004).

\bibitem{Kostelecky1989}
V. A. Kosteleck and S. Samuel, \textit{Gravitational phenomenology in higher-dimensional theories and strings.} Phy. Rev. D \textbf{40}, 1886 (1989).

\bibitem{Gomes2010}
Gomes, Marcelo, et al. \textit{Aetherlike Lorentz-breaking actions.} Phys. Rev. D \textbf{81}, 045018 (2010).

\bibitem{Bertolami2005}
O. Bertolami and J. Paramos. \textit{Vacuum solutions of a gravity model with vector-induced spontaneous Lorentz symmetry breaking.} Phys. Rev. D \textbf{72}, 044001 (2005).

\bibitem{Bertolami20051}
O. Bertolami and J. Paramos, \textit{The Flight of the bumblebee: Vacuum solutions of a gravity model with vector-induced spontaneous Lorentz symmetry breaking}, Phys. Rev. D \textbf{72}, 044001 (2005).

\bibitem{Casana2018}
R. Casana, A. Cavalcante, F.P. Poulis and E.B. Santos, \textit{Exact Schwarzschild-like solution in a bumblebee gravity model}, Phys. Rev. D \textbf{97}, 104001 (2018).

\bibitem{Chen2020}
S.B. Chen, M.Z. Wang, and J.L. Jing, \textit{Polarization effects in Kerr black hole shadow due to the coupling between photon and bumblebee field}, J. High Energy Phys. \textbf{1}, 17 (2020).

\bibitem{wang2021}
Z. Wang, S. Chen and J. Jing, \textit{Constraint on Lorentz symmetry breaking in Einstein-bumblebee theory by quasi-periodic oscillations.} preprint arXiv: 2112.02895, 2021.

\bibitem{Ovgun2019}
A. Ovgun, K. Jusufi and I. Sakall, \textit{Exact traversable wormhole solution in bumblebee gravity.} Phys. Rev. D \textbf{99}, 024042 (2019).

\bibitem{Gullu2020}
I. Gullu and A. Ovgun, (2020), \textit{Schwarzschild Like Solution with Global Monopole in Bumblebee Gravity}, arXiv:2012.02611 [gr-qc]

\bibitem{Maluf2021}
R. V. Maluf and J. C. S. Neves, \textit{Black holes with a cosmological constant in bumblebee gravity.} Phys. Rev. D \textbf{103}, 044002 (2021).

\bibitem{Ding2022}
C. Ding, X. Chen and X. Fu, \textit{ Einstein-Gauss-Bonnet gravity coupled to bumblebee field in four dimensional spacetime}. Nucl. Phys. B, \textbf{975}, 115688 (2022).

\bibitem{Ding2020}
C. Ding and C. Liu, et al. \textit{Exact Kerr-like solution and its shadow in a gravity model with spontaneous Lorentz symmetry breaking.} Eur. Phys. J. C \textbf{80}, 178 (2020).

\bibitem{Kanzi2022}
S. Kanzi,  \textit{Reply to ``Comment on `Greybody radiation and quasinormal modes of Kerr-like black hole in Bumblebee gravity model'".} Eur. Phys. J. C \textbf{82}, 1-4 (2022).

\bibitem{ReggeWheeler1957}
T. Regge and J. A. Wheeler, \textit{Stability of a Schwarzschild singularity}, Phys. Rev. \textbf{108}, 1063 (1957).

\bibitem{Zerilli1970PRL}
F. J. Zerilli, \textit{Effective potential for even parity Regge-Wheeler gravitational perturbation equations}, Phys. Rev. Lett. \textbf{24}, 13: 737 (1970).

\bibitem{Zerilli1970PRD}
F. J. Zerilli, \textit{Gravitational field of a particle falling in a schwarzschild geometry analyzed in tensor harmonics}, Phys. Rev. D \textbf{2}, 10: 2141 p(1970).

\bibitem{guage2022}
W. T. Liu, X. J. Fang, J. L. Jing and A. Z. Wang, \textit{Gauge Invariant Perturbations of General Spherically Symmetric Spacetimes}.  Sci. China-Phys. Mech. Astron. \textbf{66}, 1-14 (2023).
\bibitem{Teukolsky1973}
S. A. Teukolsky, \textit{Perturbations of a rotating black hole. I. Fundamental equations for gravitational, electromagnetic, and neutrino-field perturbations}. Astrophys. J. \textbf{185}, 635-648 (1973).

\bibitem{Rosa2012}
J. G. Rosa and S. R. Dolan, \textit{Massive vector fields on the Schwarzschild spacetime: quasinormal modes and bound states}, Phys. Rev. D  \textbf{85}, 044043 (2012).

\bibitem{PRL2018}
V. Frolov, P. Krtous and D. Kubiznak, et al. \textit{Massive vector fields in rotating black-hole spacetimes: separability and quasinormal modes.} Phys. Rev. Lett. \textbf{120}, 231103 (2018).

\bibitem{Pani2012}
P. Pani and V. Cardoso, et al. \textit{Perturbations of slowly rotating black holes: massive vector fields in the Kerr metric}, Phys. Rev. D \textbf{86}, 104017 (2012).

\bibitem{Pani2012prl}
P. Pani and V. Cardoso, et al. \textit{Black-hole bombs and photon-mass bounds}, Phys. rev. lett. \textbf{109}, 131102 (2012).

\bibitem{Pani2013prd}
P. Pani, E. Berti and L. Gualtieri, \textit{Scalar, electromagnetic, and gravitational perturbations of Kerr-Newman black holes in the slow-rotation limit}, Phys. Rev. D, \textbf{88}, 064048 (2013).

\bibitem{Pani2013prl}
P. Pani, E. Berti and L. Gualtieri, \textit{Gravitoelectromagnetic perturbations of Kerr-Newman black holes: stability and isospectrality in the slow-rotation limit}, Phys. Rev. Lett. \textbf{110}, 241103 (2013).

\bibitem{Pani2013IJMPA}
P. Pani, \textit{Advanced Methods in Black-hole Perturbation Theory} Int. J. Mod. Phys. A \textbf{28}, 22n23, 1340018 (2013).

\bibitem{Tattersall2018}
O. J. Tattersall, \textit{Kerr-(anti-) de Sitter black holes: Perturbations and quasinormal modes in the slow rotation limit}, Phys. Rev. D \textbf{98}, 104013 (2018).

\bibitem{Kojima}
Y. Kojima, \textit{Equations governing the nonradial oscillations of a slowly rotating relativistic star}, Phys. Rev. D \textbf{46}, 4289 (1992).

\bibitem{Thorne1980}
K. S. Thorne, \textit{Multipole Expansions of Gravitational Radiation} Rev. Mod. Phys. \textbf{52}, 299 (1980).

\bibitem{Thompson2016}
J. E. Thompson, B. F. Whiting and H. Chen, \textit{Gauge invariant perturbations of the Schwarzschild spacetime} Class. Quant. Grav. \textbf{34}, 174001 (2017).

\bibitem{Jing2022}
J. Jing, S. Chen, M. Sun, X. He, M. Wang and J. Wang, \textit{Self-consistent Effective-one-body theory for spinless binaries based on post-Minkowskian approximation I: Hamiltonian and decoupled equation for $\psi_B^4$} Sci. China-Phys. Mech. Astron. \textbf{65}, 260411 (2022).

\bibitem{Gerlach1979}
U. H. Gerlach and U. K. Sengupta, \textit{Gauge Invariant Perturbations On Most General Spherically Symmetric Space-times} Phys. Rev. D \textbf{19}, 2268-2272 (1979)

\bibitem{Gerlach1980}
U. H. Gerlach and U. K. Sengupta, \textit{Gauge Invariant Coupled Gravitational, Acoustical, And Electromagnetic Modes On Most General Spherical Space-times} Phys. Rev. D \textbf{22}, 1300-1312 (1980)

\bibitem{Poisson2005}
K. Martel and E. Poisson, \textit{Gravitational perturbations of the Schwarzschild spacetime: A Practical covariant and gauge-invariant formalism} Phys. Rev. D \textbf{71}, 104003 (2005).

\bibitem{Poisson2006}
B. Preston and E. Poisson, \textit{Light-cone gauge for black-hole perturbation theory} Phys. Rev. D \textbf{74}, 064010 (2006)

\bibitem{Poisson2018}
E. Corrigan and E. Poisson, \textit{EZ gauge is singular at the event horizon} Class. Quant. Grav. \textbf{35}, 137001 (2018).

\bibitem{Lin20171}
K. Lin and W.L. Qian, \textit{A matrix method for quasinormal modes: Schwarzschild black holes in asymptotically flat and (anti-) de Sitter spacetimes}. Class. Quant. Grav. \textbf{34}, 095004 (2017).

\bibitem{Lin2017}
K. Lin and W.L. Qian, et. al. \textit{A matrix method for quasinormal modes: Kerr and Kerr-Sen black holes}. Mod. Phys. Lett. A \textbf{32}, 1750134 (2017).

\bibitem{Lin2019}
K. Lin and W.L. Qian, \textit{The matrix method for black hole quasinormal modes}. Chinese Phys. C \textbf{43}, 035105 (2019).

\bibitem{Lei2021}
Y.H. Lei, M.J. Wang and J.L Jing, \textit{Maxwell perturbations in a cavity with Robin boundary conditions: two branches of modes with spectrum bifurcation on Schwarzschild black holes}. Eur. Phys. J. C \textbf{81}: 1-12 (2021).

\bibitem{Leaver1985}
E. W. Leaver, \textit{An analytic representation for the quasi-normal modes of Kerr black holes}.  Proc. Roy. Soc. Lond. \textbf{A402}, 285 (1985).

\bibitem{Konoplya2006}
R. A. Konoplya and A. V. Zhidenko, \textit{Stability and quasinormal modes of the massive scalar field around Kerr black holes}. Phys. Rev. D \textbf{73}, 124040 (2006).

\bibitem{Berti2009}
E. Berti and V. Cardoso, \textit{Quasinormal modes of black holes and black branes}. Class. Quant. Grav. \textbf{26}, 163001 (2009).

\bibitem{chen2020JHEP}
S.B. Chen, M. Wang and J. Jing, \textit{Polarization effects in Kerr black hole shadow due to the coupling between photon and bumblebee field}. J. High Energy Phys. \textbf{7}, 1-17 (2020).

\bibitem{shao2022}
R. Xu, D. Liang and L. Shao, \textit{Static spherical vacuum solutions in the bumblebee gravity model.} arXiv:2209.02209 (2022).

\bibitem{Chandbook}
S. Chandrasekhar, \textit{The Mathematical Theory of Black Holes} (Oxford University Press, Oxford, 1992).

\bibitem{isoLovelock}
C. B. Prasobh and V. C. Kuriakose, \textit{Quasinormal Modes of Lovelock Black Holes.} Eur. Phys. J. C \textbf{74}, 3136 (2014).

\bibitem{isoChSi}
S. Bhattacharyya and S. Shankaranarayanan, \textit{Distinguishing general relativity from Chern-Simons gravity using gravitational wave polarizations.} Phys. Rev. D \textbf{100}, 024022 (2019).

\bibitem{isoLoop}
D. del-Corral and J. Olmedo, \textit{Breaking of isospectrality of quasinormal modes in nonrotating loop quantum gravity black holes.} Phys. Rev. D \textbf{105}, 064053 (2022).


\end{thebibliography}
\end{document}